\documentclass[twocolumn,preprintnumbers,superscriptaddress,nofootinbib,aps,prd,floatfix]{revtex4-2}
\pdfoutput=1
\usepackage{enumerate}
\usepackage{gensymb}
\usepackage{amsmath,amssymb,dsfont}
\usepackage{amssymb}
\usepackage{mathrsfs}
\usepackage{graphicx}
\usepackage{slashed}
\usepackage{xspace,slashed}
\usepackage{hyperref}
\usepackage[dvipsnames]{xcolor}
\hypersetup{colorlinks=true, citecolor=Blue, urlcolor=Blue, linkcolor=Blue}
\usepackage[normalem]{ulem}
\usepackage{subfigure,orcidlink}
\usepackage[autostyle]{csquotes}
\usepackage{multirow,array}
\usepackage{float}
\usepackage{tabularray}
\UseTblrLibrary{booktabs}
\usepackage[absolute,overlay]{textpos} 

\begin{document}


\title{Signature of GW from the Hidden Sector in light of the Recent LZ Event}

\author{Sujit Kumar Sahoo \orcidlink{0000-0002-9014-933X}}
\email{sujitks@imsc.res.in}
\affiliation{Institute of Mathematical Sciences (IMSc), Chennai 600113, India.}

\author{Narendra Sahu \orcidlink{0000-0002-9675-0484}}
\email{nsahu@phy.iith.ac.in}
\affiliation{Department of Physics, Indian Institute of Technology Hyderabad, Kandi, Telangana-502285, India.}

\author{Vicky Singh Thounaojam}
        \email{ph22resch01004@iith.ac.in}
	\affiliation{Department of Physics, Indian Institute of Technology Hyderabad, Kandi, Telangana-502285, India.}
	
\date{\today}
\begin{abstract}
The LZ collaboration has recently reported a single nuclear-recoil event at $E_R=248\pm23_\text{stat}\pm23_\text{sys}$ keV, observed in a 2.84 tonne-year exposure with a local significance of $3.4\sigma$, pointing towards an inelastic dark matter (DM) interpretation. We explain this event within a gauged $U(1)_X$ hidden-sector model, spontaneously broken by a singlet scalar whose vacuum expectation value (VEV) gives rise to: \textit{(i)} gauge boson mass, \textit{(ii)} a small Majorana mass for the vector-like singlet fermion, the DM candidate, charged under the new symmetry, which splits the fermion into two pseudo-Dirac states (mass splitting proportional to the scalar VEV) thus resulting in a purely off-diagonal vector current, \textit{(iii)} the frequency and the amplitude of the GW signal arising from the hidden sector symmetry breaking via first-order phase transition (FOPT). The symmetry-breaking scale thus provides a multi-messenger probe connecting terrestrial direct-detection kinematics at the LZ and early-universe FOPT thermodynamics while obtaining thermal DM relic abundance. We delineate the viable parameter space compatible with the LZ event, correct relic density, and detectable GW signals in the low frequency regimes at future interferometers like $\mu {\rm ARES}$ and THEIA while satisfying existing cosmological, astrophysical, and terrestrial bounds.
\end{abstract}

\maketitle	
\noindent
\section{Introduction}
\label{sec:intro}

Dark matter (DM) remains one of the long-standing unsolved problems in physics. Despite strong astrophysical and cosmological evidence for the existence of such a non-luminous, non-baryonic component, which constitutes almost $85\%$ of the total matter content of the Universe~\cite{Planck:2018vyg}, has not been confirmed yet by any existing DM search experiments. Recently, however, the LUX-ZEPLIN (LZ) experiment reported a single DM event (LZ230616) at an unusually high recoil energy, $E_R \simeq 248 \pm 23_\text{stat} \pm 23_\text{sys}$~keV, found while searching an extended recoil window of $5.4$--$270$~keV in a $2.84$~tonne-year exposure, with a local significance of $3.4\sigma$. Because backgrounds are expected to be very small in this high-recoil regime, the event has drawn particular attention as a possible signature of inelastic DM (iDM) models~\cite{Fan:2026kxx,Freese:2026sga,Wu:2026nhi,Yin:2026jnn,Rodd:2026tyn,Smirnov:2026aqk,Nomura:2026qyq,Bisal:2026khf,Pospelov:2026ewn,Bose:2026ndd,Nguyen:2026lui,Langhoff:2026ujr,Du:2026guj, McCabe:2026crm, Unwin:2026rdp, Lou:2026idn, Su:2026rwz, DiMauro:2026ldr, Yamashita:2026ump, Chattopadhyay:2026ryw, deLima:2026shq, Visinelli:2026kgt,Borah:2026zwf,Mahapatra:2026glu,Borah:2026ris,Bandyopadhyay:2026gjw,Paul:2026okh,Barman:2026omh,Lee:2026jxl,Nagata:2026pbj,Arcadi:2026kev,Fan:2026hzw,He:2026hqz,Kumar:2026lgi,Lee:2026jxl,Baer:2026fpy,Okada:2026upm,He:2026idw,Uttayarat:2026isp,Palmisano:2026kuj,Lian:2026hpm,An:2026pkc,Cheung:2026byg,Kotlarski:2026pep,Das:2026buc,Okada:2026eol,Bose:2026szs,Elahi:2026vlm,Jeesun:2026vzo,Cabo-Almeida:2026uqw,Chauhan:2026udz,Das:2026jsq,Chanda:2026wdl,Jho:2026zzj,Jho:2026cso,Bose:2026szs,Chattopadhyay:2026ryw,Girmohanta:2026qzo,Pallis:2026jdc}. In such models, the dark sector consists of two nearly degenerate states, $\chi_1$ and $\chi_2$, separated by a small mass splitting $\delta$. The transition from the lighter to the heavier state, $\chi_1 N \rightarrow \chi_2 N$, is endothermic, where $N$ is the detector nucleus: it proceeds only if the collision supplies a minimum recoil energy sufficient to pay for $\delta$. This kinematic threshold naturally explains a signal concentrated in a narrow, high-energy recoil window, without predicting the large low recoil energy event rate that would otherwise be expected.

In particular, a generic $U(1)_X$ scenario has also been explored as an explanation of the event in \cite{Tang:2026lgu,Das:2026buc,Zhu:2026dag}. These studies examine the LZ event in the context of a portal arising from kinetic mixing, generated either at tree level or at loop level, which allows the dark photon to mediate the inelastic transition. Such scenarios can naturally account for the event while remaining consistent with existing constraints from beam-dump experiments, and collider searches. While these studies show that a dark photon portal can accommodate the LZ event, they do not address what the underlying symmetry-breaking sector implies beyond direct detection. In this paper, we address this question in a gauged $U(1)_X$ framework, where the mass splitting $\delta$, the mediator mass $m_X$ and the gauge coupling $g_X$ are not independent inputs but are tied together by a single $U(1)_X$-breaking vacuum expectation value (VEV) and by the requirement of the observed relic density. In that case, the same parameters that reproduce the LZ event also determine the dynamics of the $U(1)_X$ phase transition in the early Universe, and hence a possible stochastic gravitational wave (GW) background. We study the viable parameter space that explains the recent LZ event and gives detectable GW signal while satisfying the correct relic density constraint.

To delineate the viable parameter space, we construct a gauged $U(1)_X$ hidden-sector extension comprising a vector-like singlet fermion $\chi$, a vector boson $X$, and a singlet scalar $\phi$~\cite{Dutta:2021wbn}. The hidden sector communicates with the Standard Model (SM) through two portals: kinetic mixing between $X_\mu$ and the SM hypercharge boson $B_\mu$, and mass mixing between $\phi$ and the SM Higgs $H$. Spontaneous symmetry breaking (SSB) of $U(1)_X$, triggered by the scalar VEV $v_\phi$ controls three things simultaneously: (1) the VEV induces a small Majorana mass for $\chi$, which supplements an already-present Dirac mass term. This small Majorana mass splits the Dirac fermion into two nearly degenerate pseudo-Dirac states, $\chi_{1,2}$, and thus explains the LZ event, (2) generates  the vector mass $m_X \propto g_X v_\phi$ via Higgs mechanism, and (3) generates GW through FOPT. As the vector current of a Majorana fermion vanishes identically, the resulting $\chi$-$X$ coupling is purely off-diagonal, connecting $\chi_1$ and $\chi_2$ while forbidding any diagonal coupling. In this setup, the relic density of DM is established through its annihilation and co-annihilation channels to $X$ and $\phi$.

In the early Universe, the breaking of $U(1)_X$ may proceed through a first-order phase transition (FOPT), as has been studied extensively in $U(1)_X $ extensions of the SM~\cite{Addazi:2017gpt,Breitbach:2018ddu,Chiang:2017zbz,Feng:2026iqo,Bhattacharyya:2026esv,Pradhan:2026inq}. The vector $X$ adds new gauge degrees of freedom to the thermal plasma, and their finite-temperature loop contributions generate a cubic term in the effective potential that creates a barrier between the symmetric and broken phases. The barrier height scales with the gauge coupling $g_X$, so a strong FOPT requires a sizable $g_X$. This is precisely the coupling that is tightly constrained by the DM relic density, since $g_X$ sets the annihilation rate. A FOPT proceeds through bubble nucleation, bubble collisions, sound waves and turbulence in the plasma, which together source a stochastic GW background. For the scales of interest here, this signal is accessible to LISA \cite{LISA:2017pwj,LISACosmologyWorkingGroup:2022kbp,Caprini:2019egz}, SKA \cite{Weltman:2018zrl}, GAIA \cite{Garcia-Bellido:2021zgu}, THEIA \cite{Garcia-Bellido:2021zgu} and $\mu$ARES \cite{Sesana:2019vho}. Earlier works on $U(1)_X$ models have used this GW signal to probe the symmetry-breaking scale, the seesaw scale and hidden-sector dynamics
\cite{Addazi:2017gpt, Breitbach:2018ddu, Huang:2026qdc,Mahapatra:2026fyv}. In our model, however, $g_X$ and $v_\phi$ (or $\delta$) are simultaneously fixed by the LZ event and the relic abundance. A GW signal in this region, if predicted, would be a cosmological smoking gun for the model, complementary to laboratory searches. A systematic connection between the data-allowed parameter space and the GW prediction is lacking. In this work we address this by performing a combined Bayesian analysis with static nested sampling, using a likelihood built from the relic density and the LZ event, and then evaluating the phase transition and GW spectrum across the viable parameter space.

Beyond this cosmological probe, the model's kinetic mixing with the SM gauge sector subjects it to stringent terrestrial constraints -- from beam-dump experiments~\cite{Andreas:2012mt,Riordan:1987aw,Tsai:2019buq,NOMAD:2001eyx, Bernardi:1985ny,Davier:1989wz}, collider, fixed-target searches~\cite{LHCb:2017trq,LHCb:2019vmc,BaBar:2016sci,KLOE-2:2018kqf,NA482:2015wmo,NA62:2023qyn,BESIII:2017fwv,CMS:2019kiy,APEX:2011dww,HPS:2018xkw,A1:2011yso,Merkel:2014avp}, and precision measurements of the electron and muon anomalous magnetic moments~\cite{Bodas:2021fsy,Muong-2:2025xyk}. Additionally, the bound from Planck 2018~\cite{Planck:2018vyg} on $\Delta N_\text{eff}$ limits the MeV-scale mass of $m_X$~\cite{Ibe:2019gpv}. We confront these terrestrial, cosmological, and solar bounds against the parameter space compatible with the LZ event while obtaining the correct DM relic density, identifying the surviving region and its complementary reach at GW observatories.

The paper is organized as follows. In Section~\ref{sec:model}, we present the model and show how the off-diagonal $\chi$-$X$ current and the SM portals give rise to inelastic WIMP-nucleon scattering. In Section~\ref{sec:LZevent}, we lay out the framework for the inelastic scattering analysis of the LZ event and a combined likelihood incorporating DM relic density. In Section~\ref{sec:fopt}, we investigate the gravitational wave signature of the
$U(1)_X$ FOPT, and identify the frequency range in which it can be detected. In Section~\ref{sec:expt_constraints}, we show the surviving parameter space in the ($m_X,\epsilon$) plane against terrestrial and collider bounds. 
We conclude in Section~\ref{sec:conclusion}.

\section{A gauged $U(1)_X$ model for hidden sector}\label{sec:model}
We extend the SM gauge symmetry by a hidden-sector $U(1)_X$. The new field content, summarized in Table~\ref{tab:fields}, consists of $U(1)_X$ gauge boson $X'_\mu$, a complex scalar singlet $\Phi$ that breaks this symmetry, and a vector-like Dirac fermion $\chi$. The hidden sector connects to the SM through two portals: kinetic mixing between $X'_\mu$ and the SM hypercharge field $B_\mu$, and mass mixing between the CP-even component of $\Phi$ and the SM Higgs.
\begin{table}[ht]
\centering
\label{tab:fields}
\small
\begin{tabular}{@{}lccccc@{}}
\toprule
 & Field & $SU(3)_c$ & $SU(2)_L$ & $U(1)_Y$ & $U(1)_X$ \\
\midrule
vector & $X'_\mu$ & $\mathbf{1}$ & $\mathbf{1}$ & $0$ & gauge\\
scalar & $\Phi$     & $\mathbf{1}$ & $\mathbf{1}$ & $0$ & $-2$ \\
fermion  & $\chi$  & $\mathbf{1}$ & $\mathbf{1}$ & $0$ & $+1$\\
\bottomrule
\end{tabular}
\caption{New field content and quantum numbers. All SM fields carry $Q_X=0$.}
\end{table}

The relevant Lagrangian term is expressed as:
\begin{align}\label{eq:Lag1}
    \mathcal{L}\supseteq &\overline{\chi}\gamma^\mu iD_\mu\chi + (D^\mu \Phi)^\dagger(D_\mu \Phi)-\tfrac{1}{4}X'_{\mu\nu}X'^{\mu\nu}-\tfrac{\epsilon}{2}X'^{\mu\nu}B_{\mu\nu}\nonumber\\
    &-\frac12 (y_L \Phi \overline{\chi_L^C} \chi_L
		+ y_R \Phi \overline{\chi_R^C} \chi_R + {\rm h.c.})\nonumber\\
    &\quad -M_D(\overline{\chi_L}\chi_R+\overline{\chi_R}\chi_L) -V(H,\Phi),
\end{align}
where $D_\mu = \partial_\mu - iQ_Xg_XX'_\mu$ is the $U(1)_X$ covariant derivative and $X'_{\mu\nu} = \partial_\mu X'_\nu - \partial_\nu X'_\mu$ is the corresponding field strength tensor; $\epsilon$ denotes the dimensionless kinetic mixing parameter. We note that the Majorana Yukawa term is retained in the Lagrangian, as it remains invariant under the gauge symmetry considered here. The scalar potential is given by,
\begin{align}
    V(H,\Phi) = {}& \lambda_h\left(H^\dagger H-\tfrac{v_h^2}{2}\right)^2
		+\lambda_\phi\left(\Phi^*\Phi-\tfrac{v_\phi^2}{2}\right)^2 \nonumber\\
		&+\lambda_{h\phi}\left(H^\dagger H-\tfrac{v_h^2}{2}\right)\left(\Phi^*\Phi-\tfrac{v_\phi^2}{2}\right),
		\label{eq:potential}
\end{align}
where $v_h$ and $v_\phi$ denote the VEV of Higgs and $\Phi$ field, respectively. Expanding the scalar fields with $H=(h+v_h)/\sqrt{2}$ and $\Phi=(\phi+v_\phi)/\sqrt{2}$ to quadratic order gives the CP-even mass matrix
\begin{equation}
M^2_{hh}=2\lambda_h v_h^2, \quad
M^2_{\phi}=2\lambda_\phi v_\phi^2, \quad
M^2_{h\phi}=\lambda_{h\phi}\,v_h\,v_\phi .
\end{equation}
Diagonalizing via mixing angle $\alpha$,
\begin{equation}
h=\cos\xi\, h_1-\sin\xi\, h_2, \qquad
\phi=\sin\xi\, h_1+\cos\xi\, h_2,
\label{eq:scalarmix}
\end{equation}
with
\begin{equation}
\tan2\xi=\frac{2M^2_{h\phi}}{M^2_{hh}-M^2_{\phi\phi}} .
\end{equation}
In the small $\tan\xi\ll 1$ limit, the masses of the physical states ($h_1,h_2$) can be approaximated with $m_h\simeq 2\lambda_H v_h^2$ and $m_\phi\simeq 2\lambda_\phi v_\phi^2$, respectively.
Using the above relations, the scalar quartic couplings can be written as:
\begin{align}
\lambda_h &= \frac{\cos^2\xi\, m_h^2+\sin^2\!\xi\, m_\phi^2}{2v_h^2}, \\
\lambda_\phi &= \frac{\sin^2\xi\, m_h^2+\cos^2\!\xi\, m_\phi^2}{2v_\phi^2}, \\
\lambda_{h\phi} &= \frac{\sin\xi\cos\xi\,(m_h^2-m_\phi^2)}{v_hv_\phi}.
\label{eq:lamHS}
\end{align}
Through the SSB in the hidden sector, the respective gauge boson acquires a mass $m_X=2g_X v_\phi$. Similarly, the fermions acquire Majorana mass via Yukawa term in the Lagrangian given in Eq.~\eqref{eq:Lag1}. Thus, the fermion mass Lagrangian becomes:
\begin{align}
   -\mathcal{L}_{\text{mass}}&\supseteq \tfrac{1}{2}(M_L \overline{\chi_L^C}\chi_L + M_R \overline{\chi_R^C}\chi_R+\text{h.c}) \nonumber\\
   & \quad + M_D(\overline{\chi_L}\chi_R+\overline{\chi_R}\chi_L), 
\end{align}
with $M_L=y_Lv_\phi/\sqrt2$, $M_R=y_Rv_\phi/\sqrt2$, and the symmetric mass matrix
\begin{equation}
	\mathcal{M}=\begin{pmatrix} M_L & M_D \\ M_D & M_R \end{pmatrix}.
\end{equation}
The mass matrix can be diagonalized by applying a $2\times2$ orthogonal rotation matrix $O$. However, in the pseudo-Dirac regime, $M_D\gg M_L,M_R$, which is the parameter region relevant for our work. In this limit, the diagonalization gives the smaller eigenvalue with a negative sign. This sign does not imply a negative physical mass; rather, it indicates a relative phase associated with the corresponding Majorana eigenstate. To express the physical masses in the conventional positive-mass basis, this phase can be absorbed into the definition of the corresponding mass eigenstate.

To account for this phase, we redefine the rotation matrix as $U=O\rho$, where the matrix $\rho={\rm diag}(\rho_1,\rho_2)$ contains the phases, with $\rho_1^2=-1$ and $\rho_2^2=1$. This phase choice removes the unphysical sign of the corresponding mass eigenvalue while preserving the physical content of the theory. The resulting field transformation is
\begin{equation}
\begin{pmatrix}
X_L \\
X_R^c
\end{pmatrix}=
\begin{pmatrix}
    i\cos\theta & \sin\theta \\
-i\sin\theta & \cos\theta
\end{pmatrix}
\begin{pmatrix}
    \chi_1 \\
    \chi_2
\end{pmatrix},
\end{equation}
where $\chi_1$ and $\chi_2$ denote the physical mass eigenstates. In the pseudo-Dirac limit, their masses are therefore given by
\begin{equation}
	\begin{split}
	    m_{1} &\simeq M_D-\frac{M_L+M_R}{2},\\
        m_{2} &\simeq M_D+\frac{M_L+M_R}{2}.
	\end{split}
\end{equation}
Thus, the mass splitting is given by 
\begin{equation}
    \delta \equiv m_{2}-m_{1} \simeq M_L+M_R .
	\label{eq:splitting}
\end{equation}
The two states are quasi-degenerate, differing only by the small Majorana mass splitting, with the mixing angle taking its maximal value $\theta \to \pi/4$ (corresponding to $y_L = y_R \equiv y$); the splitting is then given by $\delta = y v_\phi/\sqrt{2}$.

We identify the lightest fermion $\chi_1$ as the dark matter candidate and its stability is ensured by the residual $\mathcal{Z}_2$ symmetry.
Now expanding the fermion interaction term Eq.~\eqref{eq:Lag1} in Majorana mass basis, the diagonal vector current vanishes identically, and we get the off-diagonal vector current as:
\begin{equation}
	\mathcal{L}_{\rm int}\supseteq{i g_X}\,X'_\mu
	(\bar\chi_1\gamma^\mu\chi_2).
	\label{eq:offdiag}
\end{equation}
$\chi_1$ and $\chi_2$ being two pseudo-Dirac states has no tree-level vector coupling with $X'_\mu$, only an inelastic transition from one state to another with mass splitting $\delta$.

Since we consider a gauge kinetic mixing for the portal, we follow the conventions presented in \cite{Borah:2025cqj}. Following the rotation from the unprimed basis to the mass eienstate,
\begin{equation*}
    \begin{pmatrix} Z'_\mu \\ X'_\mu \end{pmatrix}=
\begin{pmatrix}
\cos\zeta & -\sin\zeta \\
\sin\zeta & \cos\zeta
\end{pmatrix}
\begin{pmatrix} Z_\mu \\ X_\mu \end{pmatrix}.
\end{equation*}
The mixing angle is given by 
\begin{equation}
    {\rm tan}(2\zeta)\simeq \frac{2\epsilon s_W}{1-\frac{m_X^2}{m_Z^2}},
\end{equation}
where $s_W \equiv \sin\theta_W$, and $m_Z$ denotes the mass of the neutral gauge boson in the SM. Then, the interaction Lagrangian of the new gauge field $X_\mu$ with the SM fermion is given by
\begin{align}\label{eq:Xcoupling}
    \mathcal{L}_{\mathrm{int}}&\supseteq \sum_f \left[X_\mu \bar{f}\gamma^\mu
\left(g_{V,X}^f-g_{A,X}^f\gamma^5\right)f \right]\nonumber\\
&\quad - i\frac{\epsilon s_Wm_Z^2}{m_Z^2-m_X^2} Z_\mu g_X \overline{\chi_1}\gamma^\mu\chi_2 -  iX_\mu g_X \overline{\chi_1}\gamma^\mu\chi_2
\end{align}
where the couplings are simplified assuming $\zeta$ is very small, giving roughly
\begin{equation}\label{eq:Xf_couplings}
\begin{split}
g_{V,X}^f
&\simeq
-\epsilon e c_W Q_f
-\frac{g\epsilon s_W}{2c_W}
\frac{m_X^2}{m_Z^2-m_X^2}
\left(T_3^f-2s_W^2Q_f\right),
\\
g_{A,X}^f
&\simeq
-\frac{g\epsilon s_W}{2c_W}
\frac{m_X^2}{m_Z^2-m_X^2}\,T_3^f.
\end{split}
\end{equation}

where $\quad c_W \equiv \cos\theta_W, \quad e = g s_W.$ and $T_3^f$ denotes the third component of the weak isospin and $Q_f$ is the electric charge of the SM fermions. The Lagrangian dictated by Eq.~\eqref{eq:Xcoupling} shows that there can be two different types of interactions as follows
\begin{align*}
    &{\rm VV}: \quad (\overline{\chi_2}\gamma_\mu
\chi_1)(\overline{N}\gamma^\mu N)\\
&{\rm VA}: \quad (\overline{\chi_2}\gamma_\mu
\chi_1)(\overline{N}\gamma^\mu \gamma^5 N).
\end{align*}
The leading order non-relativistic effective field theory (NREFT) operator relevant for the vector-vector interaction is \cite{Anand:2013yka,Fitzpatrick:2012ix,Barello:2014uda} $\mathcal{O}_1=\mathds{1}_\chi\mathds{1}_N$, with contributions arising from the electromagnetic and weak current. Whereas, for the vector-axial part, the equivalent NREFT operator is a combination of $\mathcal{O}_7,\,\mathcal{O}_9$ and $\mathcal{O}_{10}$ \cite{Barello:2014uda}, given by 
\begin{equation*}
    {\rm VA} \equiv \quad -2(\mathcal{O}_7-\frac{m_N}{m_{\chi}}\mathcal{O}_9+i\frac{m_N\delta}{|\mathbf{q}|^2}\mathcal{O}_{10}).
\end{equation*}
Here, $m_N$ denotes the mass of the nucleon. The corresponding coefficients are derived from the weak current, which is heavily suppressed by the mass of the $Z$ boson. In contrast, the vector coupling receives an additional electromagnetic contribution (see Eq.~\eqref{eq:Xf_couplings}) on top of the weak-current part. This coupling appears only in the $\mathcal{O}_1$ operator and therefore remains unsuppressed. Consequently, the leading-order operator for inelastic DM scattering in our analysis is $\mathcal{O}_1$, with the relevant coefficients calculated in Appendix~\ref{app:nreft}.


\section{inelastic DM and LZ event}\label{sec:LZevent}
In our framework, DM consists of two nearly degenerate states, $\chi_1$ and $\chi_2$, with masses $m_1<m_2$ and splitting $\delta \equiv m_2-m_1$. The dominant interaction is off-diagonal, $\chi_1 N \to \chi_2 N$ (see Eq.\eqref{eq:offdiag}). In such models the DM upscatters to a state heavier by $\delta$ during the collision. This raises the velocity required to produce a given recoil and removes the low-energy part of the spectrum. The minimum DM speed needed to produce a recoil $E_R$ on a nucleus of mass $m_A$ is given by \cite{Barello:2014uda,Tucker-Smith:2001myb}
\begin{equation}
    v_{\rm min} (E_R)=\frac{1}{\sqrt{2 m_A E_R}}\left ( \frac{m_A E_R}{\mu}+\delta\right)
    \label{eq:v_min}
\end{equation}
where $\mu$ is the DM–nucleus reduced mass. For $\delta>0$ (endothermic scattering) a kinematic threshold appears, and only the high-velocity tail of the halo can excite $\chi_2$. We model the Galactic dark matter velocities using a Maxwellian
distribution truncated at the Galactic escape speed. In the Earth's
rest frame, the normalized velocity distribution is
\begin{equation}
\begin{split}
f_{\rm lab}(\mathbf{v})
={}&N\left(\frac{3}{2\pi v_{\rm rms}^{2}}\right)^{3/2}
\exp\left[
-\frac{3|\mathbf{v}+\mathbf{v}_{E,\rm Gal}|^{2}}
{2v_{\rm rms}^{2}}
\right]\times\\
&\qquad\Theta\left(
v_{\rm esc,Gal}
-|\mathbf{v}+\mathbf{v}_{E,\rm Gal}|
\right),
\end{split}
\label{eq:lab_velocity_distribution}
\end{equation}
where $\mathbf{v}$ is the dark matter velocity relative to the Earth, $v_{\rm rms}=\sqrt{{3/2}}|\mathbf{v}_{\rm rot,Gal}|$ is the root-mean-square speed of the underlying untruncated Galactic Maxwellian, $v_{\rm esc,Gal}$ is the Galactic escape speed, and $\Theta$ is the Heaviside step function. The normalization factor is
\begin{equation}
N=
\left[
\operatorname{erf}(z)
-\frac{2z}{\sqrt{\pi}}e^{-z^{2}}
\right]^{-1},
\qquad
z=\sqrt{\frac{3}{2}}\,
\frac{v_{\rm esc,Gal}}{v_{\rm rms}},
\label{eq:velocity_normalization}
\end{equation}
such that $\int d^{3}v\,f_{\rm lab}(\mathbf{v})=1$. We neglect the Earth's orbital velocity and hence the annual modulation of the rate. The Earth's Galactic velocity is therefore approximated by the Sun's Galactic velocity, $\mathbf{v}_{E,\rm Gal}
\simeq \mathbf{v}_{\odot,\rm Gal} = \mathbf{v}_{\rm rot,Gal} +\mathbf{v}_{\odot,\rm pec}$,
where $|\mathbf{v}_{\rm rot,Gal}|=220~\mathrm{km\,s^{-1}}$ and $|\mathbf{v}_{\odot,\rm pec}|= 18.04~\mathrm{km\,s^{-1}}$ \cite{Sch_nrich_2010}. We adopt the halo parameters in the range $498\leq v_{\rm esc,Gal} \leq 608~\mathrm{km\,s^{-1}}$ \cite{Smith:2006ym}. The maximum velocity of the DM therefore is given by $v_{\rm max}=608~{\rm kms^{-1}}$ , and this sets the largest kinematically accessible mass splitting, for endothermic scattering, given by
\begin{equation}
    \delta_{\max}=\tfrac12\mu v_{\max}^2.
    \label{eq:delta_max}
\end{equation}

\begin{figure}[t]
    \centering
    \includegraphics[width=0.9\linewidth]{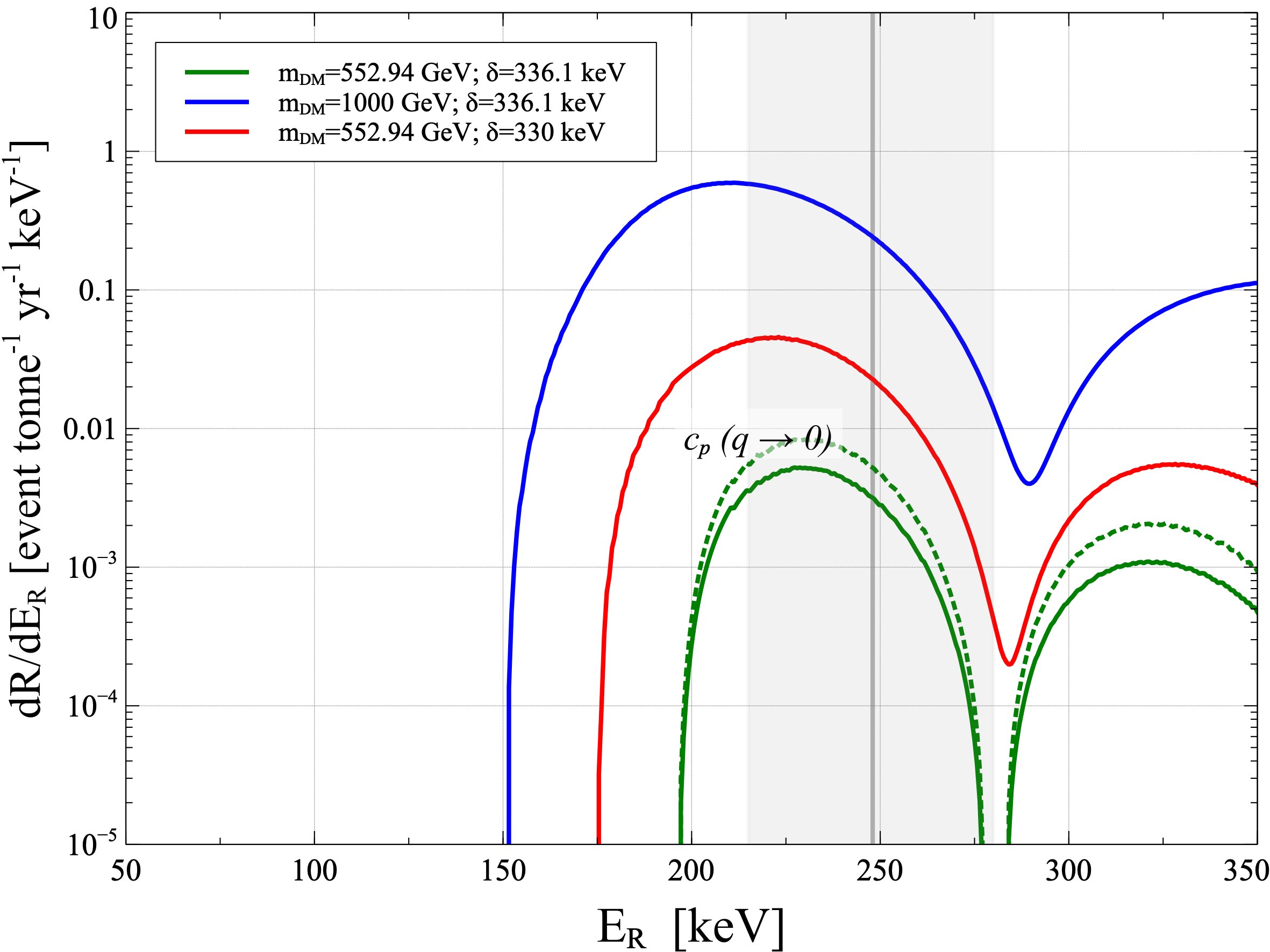}
    \caption{The predicted differential event rate as a function of the recoil energy $E_R$ for $g_X=0.49$ and $\epsilon=4.7\times10^{-5}$. The three solid curves correspond to a fixed mediator mass $m_X=0.469$ GeV, with the DM mass and mass splitting $\delta$ given in the legend. The resulting WIMP-nucleon cross-section is $8.6\times 10^{-38}$ ${\rm cm }^2$. For comparison purpose, we provide the green dotted curve to show the effect of coefficient $c_p$ of the NREFT operator $\mathcal{O}_1$ in the zero momentum transfer limit. The gray line and shaded region indicate the recently reported single-event result from LZ \cite{LZ:2026axp}.}
    \label{fig:LZ_spectrum}
\end{figure}

In this model, the vector-vector interaction between the DM current and the nucleon current reduces, at leading order in the non-relativistic expansion, to the NREFT operator $\mathcal{O}_1=\mathds{1}_\chi\mathds{1}_N$. In the kinetic-mixing scenario, and at leading order in the electromagnetic charge, this interaction is strongly proton dominated (see Appendix~\ref{app:nreft}). The corresponding isospin-basis coefficients are $c^0=c_1^p+c_1^n$ and $c^1=c_1^p-c_1^n$, with $c_1^p$ and $c_1^n$ given by
\begin{equation}
    c_1^p(q) \simeq \frac{g_X \epsilon e {\rm ~c_W}}{q^2+m_X^2};\quad c_1^n \simeq 0
    \label{eq:coefficeints}
\end{equation}
where $q=|\mathbf q|$ is the momentum transfer. To compute the event rate, we use the publicly available package, \texttt{WimPyDD} \cite{Jeong:2021bpl}. The coefficients are supplied as functions of $q$ in the isospin basis. For each xenon isotope and recoil energy the momentum transfer is fixed by $q=\sqrt{2m_AE_R}$, and the coefficients are evaluated at that value. The differential rate is then obtained by summing over isotopes $A$ weighted by their abundances,
\begin{equation}\label{eq:event_rate}
    \frac{dR}{dE_R}=\frac{\rho_{\rm DM}}{m_\text{DM}}\sum_AN_A\int_{v>v_{min}}\frac{d\sigma_A}{dE_R}v\,f_{lab}(\mathbf{v},t)d^3v,
\end{equation}
where $N_A$ denotes the number of target nuclei per unit detector mass. The differential cross-section is given by \cite{Jeong:2021bpl} 
\begin{equation}
    \frac{d\sigma_A}{dE_R}=\frac{2m_A}{4\pi}\frac{1}{v^2}\Big[\frac{1}{2j_{\rm DM}+1}\frac{1}{2j_A+1}\sum_{\rm spins}|\mathcal M_A|^2 \Big],
\end{equation}
and is shown as an average over the initial WIMP and target spins. Here, $m_A$ denotes the mass of the Xenon isotope. In non relativistic effective theory, the amplitude $|\mathcal{M}_A|$ is calculated by combining WIMP and nuclear responses \cite{Anand:2013yka,Fitzpatrick:2012ix,Jeong:2021bpl}. Fig.~\ref{fig:LZ_spectrum} shows the resulting recoil spectra. The spectra exhibit a sharp kinematic onset at low $E_R$, below which endothermic scattering is forbidden. For fixed $m_{\rm DM}=552.94$~GeV, lowering the mass splitting from $\delta=336.1$ to $330$~keV (green to red) shifts the onset toward lower recoil energies and substantially enhances the scattering rate. Similarly, increasing the DM mass from $552.94$ to $1000$~GeV (green to blue) shifts the onset to lower $E_R$ and leads to a much larger rate, reflecting the reduced kinematic suppression for a larger DM--nucleus reduced mass. The solid and dotted green curves compare the momentum-dependent and momentum-independent choices of $c_p$, respectively, for $m_X=469$~MeV. Since the mediator mass is comparable to the momentum transfer in the relevant recoil-energy range, the finite-mass propagator suppresses the rate relative to the contact-interaction approximation, resulting in a larger rate for the momentum-independent case.

\subsection{Setup for our analysis}
Following the two-bin analysis presented in \cite{Su:2026rwz}, we construct the likelihood by defining two non-overlapping recoil-energy bins: a low-energy bin, $E_1 \in [100,200]$ keV, and a high-energy bin, $E_2 \in [200,271]$ keV. We require the observed event count in $E_1$ to be null, while the count in $E_2$ is required to be consistent with one event. This construction reproduces the observed configuration, namely a single event in the high-energy bin with no accompanying events at lower recoil energy. Since the two bins are non-overlapping, they can be treated as statistically independent Poisson counting experiments, and the total likelihood factorizes as a product of the individual Poisson likelihoods. Neglecting background and its uncertainties, for each bin,  we write down the likelihood function as
\begin{equation}
    \mathcal{L}_{\rm LZ}(\Theta) = \prod_{i=1}^{2} \frac{\mathcal{N}_i(\Theta)^{\,n_i}\, e^{-\mathcal{N}_i(\Theta)}}{n_i!},
    \label{eq:likelihood}
\end{equation}
where $(n_1,n_2) = (0,1)$ are the observed event counts in the two bins, and $\mathcal{N}_i(\Theta)$ denotes the corresponding theoretical event count, obtained by integrating the differential recoil-energy spectrum $dR/dE_R$, weighted by the detector efficiency $\eta(E_R)$ and the exposure $\mathcal{E}=2.84$ tonne-yr \cite{LZ:2026axp}, over the energy range of bin $i$, namely
\begin{equation}
    \mathcal{N}_i(\Theta)=\mathcal{E}\int_{\rm bin~i}dE_R~\epsilon(E_R)\frac{dR}{dE_R}.
\end{equation}

While the LZ likelihood of Eq.~\eqref{eq:likelihood} depends on the full parameter set $\Theta=\{m_{\rm DM},~\delta,~m_X,\,\epsilon,\,g_X\}$, the relic abundance $\Omega h^2$ is sensitive to only a subset of these. Since $m_X \ll m_{\rm DM}$ in the region of interest, the dark matter annihilates through the channel, $\chi_i\chi_j\to XX, X\phi$. The process $\chi_i\chi_j\to X\phi$ is sub-dominant due to the involvement of the $\delta$, whereas $\chi_i\chi_j\to XX$ is the dominant channel. Therefore, the relic abundance depends predominantly on $g_X$ and $m_{\rm DM}$, although we included both the processes in our numerical analysis. We choose three benchmark points, $m_\phi = (0.01,0.1,1)$ GeV, for separate numerical studies, and examine in Section~~\ref{sec:fopt} how this choice correlates the LZ and relic density favored parameter space with the gravitational-wave signatures expected from a first-order phase transition (FOPT.

The same coupling $g_X$ that sets the annihilation rate also controls the size of the nucleon transition amplitude relevant for scattering, so the relic abundance and the direct-detection rate are not independent observables. This connection is sharpened by the kinematics of the LZ event itself: the momentum transfer at the candidate recoil, $E_R=248$ keV on a xenon target, is $q=\sqrt{2m_A E_R}\simeq246$ MeV, comparable to the GeV-scale mediator masses preferred by the fit. When $m_X$ lies close to this scale, the full mediator propagator $1/(q^2+m_X^2)$ must be retained rather than its heavy-mediator, contact-operator limit, which both reshapes the recoil spectrum $dR/dE_R$ and enhances the scattering rate relative to a TeV-scale mediator by many orders of magnitude. A single-observable fit to the LZ event alone can therefore favor a region of $(m_{\rm DM},\delta,m_X,\epsilon,g_X)$ that may be inconsistent with the observed relic density, and vice versa. A combined likelihood analysis is therefore necessary to identify the region of parameter space in which both observables are satisfied simultaneously, while giving observable GW.

We implement this by defining the total likelihood as the product of the LZ likelihood and a relic-density likelihood
\begin{equation}
    \mathcal{L}_{\rm tot} = \mathcal{L}_{\rm LZ}\times\mathcal{L}_\Omega, \qquad
\ln\mathcal{L}_\Omega = -\frac{1}{2}\left(\frac{\Omega h^2(\Theta)-0.1200}{\sigma_\Omega}\right)^2,
\label{eq:combined_likelihood}
\end{equation}
where $\Omega h^2(\Theta)$ is the relic abundance predicted at a given point in parameter space and the central value $0.1200$ is the Planck measurement~\cite{Planck:2018vyg}. The variance combines the reported Planck uncertainty, $\sigma_{\rm Planck}=0.0012$, in quadrature with a $10\%$ theoretical uncertainty on the predicted abundance, $\sigma_\Omega=\sqrt{\sigma_{\rm Planck}^2+(0.1\,\Omega_{\rm DM} h^2)^2}$, following the standard treatment adopted in relic-density global fits~\cite{Strege:2014ija} to account for numerical uncertainties in the freeze-out calculation. A Gaussian form is appropriate here because the relic density is a continuous, precisely measured quantity with a well-defined central value and a symmetric uncertainty, in contrast to the LZ event count, which is discrete, small-number data and is more properly described by the Poisson likelihood of Eq.~\eqref{eq:likelihood}. 

\begin{figure}[t]
    \centering
    \includegraphics[width=0.9\linewidth]{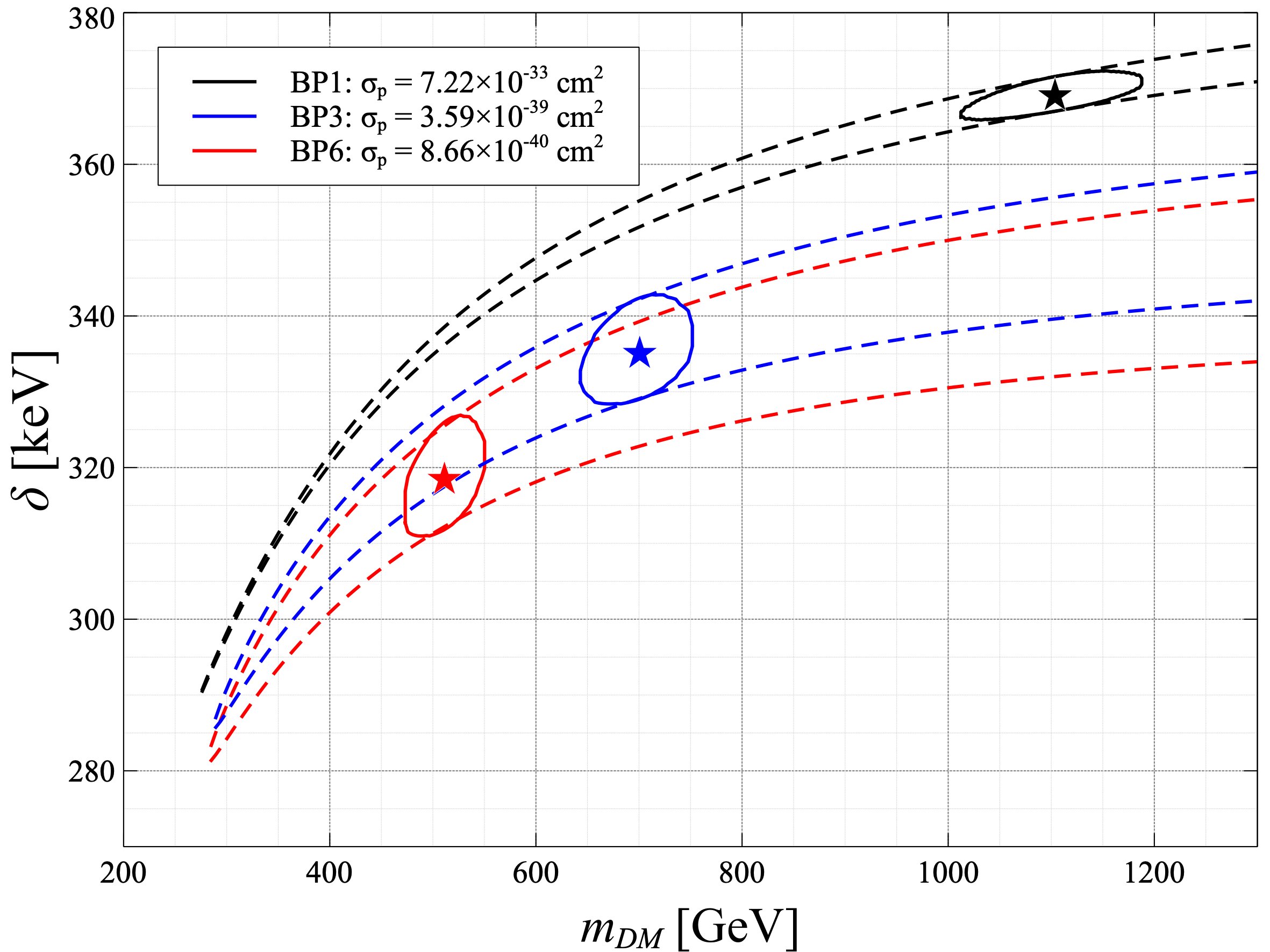}
    \caption{$1\sigma$ contours in the plane of  ($m_{\rm DM},\delta$) for the recent LZ event at $E_{R}=248\pm 23_{\rm stat} \pm 23_{\rm sys}~ ~\text{keV}$ and the observed relic density, are shown for BP1 (black), BP3 (blue) and BP6 (red). While the dashed contours satisfy only the LZ, the solid contours satisfy both LZ event as well as the observed relic density. The stars mark shows the best fit for the combined analysis. The full parameter values for BP1, BP3 and BP6 are listed in Table~\ref{tab:FOPT_BPs}.}
    \label{fig:likelihood}
\end{figure}

Fig.~\ref{fig:likelihood} shows the $1\sigma$ region in the ($m_{\rm DM},\delta$)
plane for three benchmark points, BP1 (black), BP3 (blue) and BP6 (red) given in Table~\ref{tab:FOPT_BPs} in Section~\ref{sec:fopt}. The region is
defined by $\Delta\chi^2 \equiv -2\left[\ln\mathcal{L}_{\rm LZ}-\ln\mathcal{L}_{\rm max}\right]\leq 2.30$. The dashed contours show the preferred parameter space considering only the LZ likelihood analysis indpendent of the DM relic constraints. When the DM relic density constraint is imposed, the parameter $g_X$ enforces the range of the DM mass to lie within a very narrow region and hence the combined likelihood shrinks the allowed region to a small subset, as shown by the solid contours. The inelastic DM-proton cross section is $\sigma_p=\frac{\mu_{\chi p}^2}{\pi}\,|c^p_1|^2$, where $\mu_{\chi p}$ is the DM-proton reduced mass and $c_1^p=e\,\epsilon\,g_X\,c_{\rm W}/m_X^2$ is the effective coupling induced by $X$ exchange through kinetic mixing.
The relatively large $\sigma_p\sim 10^{-33}~\text{cm}^2$ is allowed for two reasons. First, because $\delta$ lies close to the kinematic threshold for scattering on xenon, only the high-velocity tail of the halo contributes, so a large cross section is needed to obtain $\mathcal{O}(1)$ events. Second, the $\chi$-$X$ coupling is purely off-diagonal and therefore gives no elastic scattering. The only elastic channel is $t$-channel exchange of $\phi$ and the Higgs through their mixing, which is constrained by the elastic direct-detection limits. This channel is controlled by the $\phi$-$h$ mixing angle and is suppressed for a sufficiently small mixing angle. The strong elastic bounds are thereby evaded without affecting the inelastic signal.

\subsection{Numerical Analysis}
We sample the five-dimensional parameter space $\Theta = \{m_{\rm DM},\,\delta,\,m_X,\,\epsilon,\,g_X\}$ using the nested-sampling algorithm implemented in the \texttt{dynesty} package~\cite{Higson:2018cwj}. Nested sampling is well suited to this problem because it simultaneously produces posterior samples $\mathcal{P}(\Theta|\rm D)=\mathcal{L}_{\rm tot} ({\rm D}|\Theta).\pi(\Theta)/\mathcal{Z}$ and an estimate of the Bayesian evidence, $\mathcal{Z}=\int\mathcal{L}_{\rm tot} (\Theta)\pi(\Theta)d\Theta$ while remaining robust to the multimodal and strongly curved likelihood surfaces that arise here from the sharp kinematic threshold on $\delta$ and the sensitivity of the direct-detection rate to $m_X$ near the momentum transfer scale of the LZ event. Here, the prior is denoted by $\pi(\Theta)$ and its volume is bounded as
\begin{equation}
    \begin{aligned}
    100~{\rm GeV} &\le m_{\rm DM} \le 20~{\rm TeV}, \\
    50~{\rm keV} &\le \delta \le 600~{\rm keV}, \\
    10^{-3}~{\rm GeV} &\le m_X \le 90~{\rm GeV}, \\
    10^{-10} &\le \epsilon \le 10^{-2}, \\
    0.01 &\le g_X \le 2.0.
\end{aligned}
\label{eq:prior}
\end{equation}
The upper bound on $\delta$ is set comfortably above the kinematic ceiling for our heaviest sampled DM mass, so the sampler is free to discover the ceiling itself rather than having it imposed as a hard prior edge. The range of $m_X$ brackets the momentum transfer of the LZ event of interest, $q\simeq246$ MeV, allowing the sampler to explore both the light-mediator regime relevant to the endothermic signal and the decoupling limit\footnote{In the most extreme case for heavy-mediator - contact-operator approximation, the rate becomes so suppressed that essentially no signal is produced at any reasonable $\epsilon$ — the mediator has effectively ``decoupled" from the direct-detection observable.} at larger $m_X$.

All parameters except the mass splitting $\delta$ are sampled log-uniformly, reflecting the expectation that the dark coupling $g_X$, the kinetic mixing $\epsilon$, the mediator mass $m_X$, and the DM mass can plausibly span several orders of magnitude. At each sampled point $\Theta$, the total log-likelihood is evaluated as the sum of the LZ contribution and the relic-density contribution described in Eq.~\eqref{eq:combined_likelihood}. The LZ term is computed analytically from the tabulated differential rate and detector response using the package \texttt{WimPyDD} \cite{Jeong:2021bpl}. The relic abundance $\Omega h^2(\Theta)$, however, cannot be obtained in closed form for this model, and is instead computed numerically at each parameter point using \texttt{micrOMEGAs}~\cite{Belanger:2026asz}

\begin{figure*}[t]
    \centering
    \includegraphics[width=1.0\linewidth]{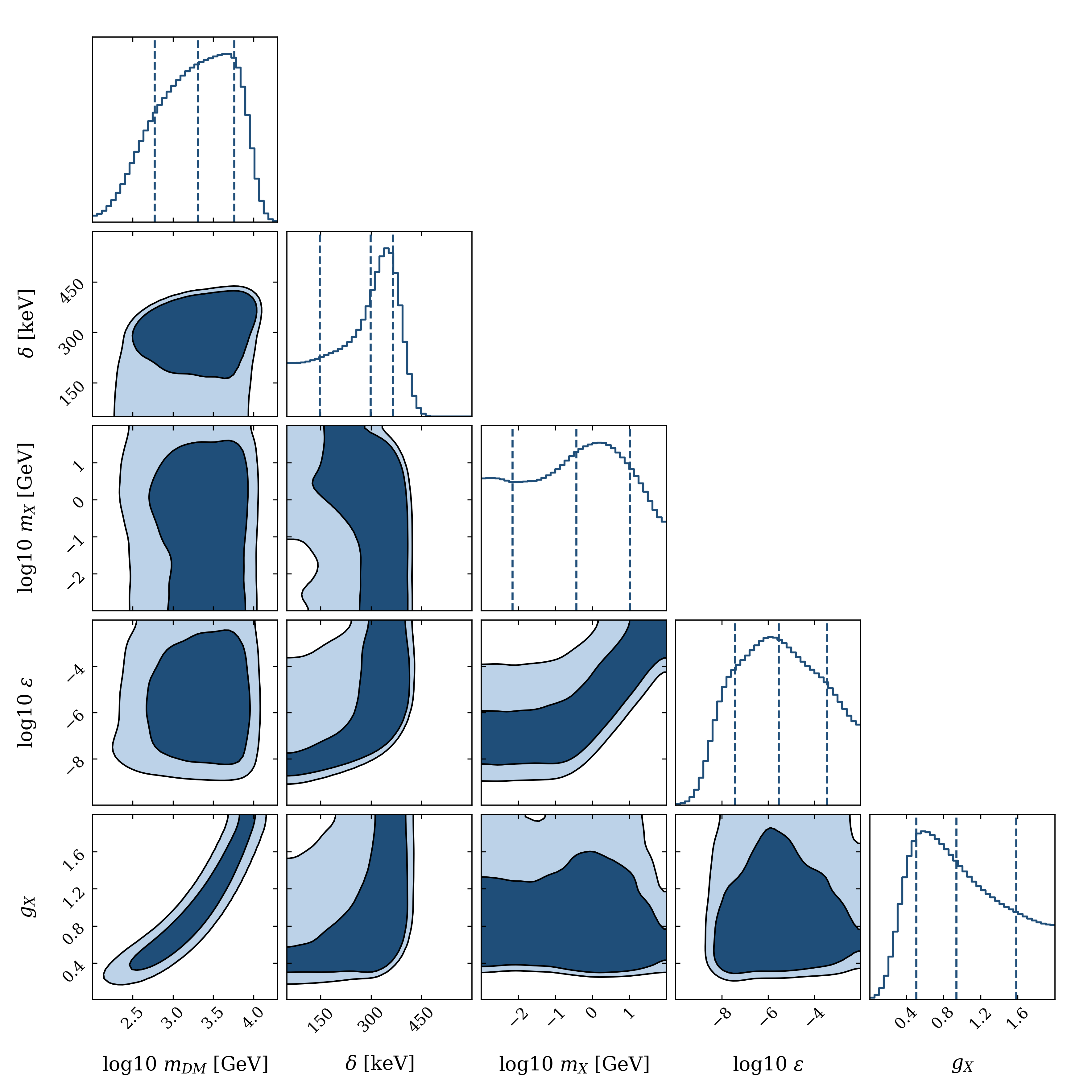}
    \caption{Corner plot of the weighted posterior distribution from the combined LZ + relic-density nested-sampling analysis, in the five-dimensional parameter space $\Theta=(m_{\rm DM},\,\delta,\,m_X,\,\epsilon,\,g_X)$. Diagonal panels show the 1D marginal posterior for each parameter, with vertical dashed lines marking the 16th, 50th, and 84th percentiles. Off-diagonal panels show the corresponding 2D marginals, with the dark- and light-shaded regions denoting the 68\% and 95\% credible regions, respectively. $m_{DM}$, $m_X$, and the kinetic mixing $\epsilon$ are shown on a logarithmic axis; the mass splitting $\delta$ and the dark gauge coupling $g_X$ are shown linearly.}
    \label{fig:weighted posterior}
\end{figure*}

The posterior distribution of the five model parameters is shown in Fig.~\ref{fig:weighted posterior}. The one- and two-dimensional distributions are obtained by marginalizing the weighted joint posterior over the remaining parameters, using the priors specified in Eq.~\eqref{eq:prior}. From Fig.~\ref{fig:weighted posterior}, we see that:
\begin{enumerate}
  \item the marginal posterior for DM masss rises slowly from a few hundred GeV, turns over in the multi-TeV range, and falls off around $20~\mathrm{TeV}$, with a median of $m_{\rm DM}=2.04^{+3.71}_{-1.44}~\mathrm{TeV}$ ($\log_{10}(m_{\rm DM}/\mathrm{GeV})=3.31^{+0.45}_{-0.54}$). The sharp fall in the high mass region is due to the $1/m_{\rm DM}$ suppression in the event rate (see Eq.~\eqref{eq:event_rate}), even though it is allowed by the relic density. On the other hand, the DM mass is weakly constrained from the lower mass side.
  \item The mass splitting has a much sharper, asymmetric posterior, with median $\delta=298^{+66}_{-151}~\mathrm{keV}$ and a mode at somewhat larger values, close to the upper edge. The sharp upper edge reflects the kinematic ceiling: for $v_{\max}\simeq 608~\mathrm{km/s}$ and $\mu\to m_A$, it saturates at $\simeq 410~\mathrm{keV}$. The smaller $\delta$ down to $50~{\rm keV}$ are allowed because one can choose smaller cross-section $\sigma$ in order to produce one event. We note that $\delta$ has no role in the relic density estimation.
  \item The mediator mass and kinetic mixing are individually weakly constrained. The posterior of $\log_{10}(m_X/\mathrm{GeV})$ has a broad maximum near $1~\mathrm{GeV}$ and a plateau toward the MeV scale, with median $m_X=0.37^{+9.94}_{-0.36}~\mathrm{GeV}$. The kinetic mixing has median $\epsilon=(2.9^{+345}_{-2.8})\times10^{-6}$, and its $68\%$ interval, $3.7\times10^{-8}$ to $3.5\times10^{-4}$, spans about four orders of magnitude. The $(m_X,\epsilon)$ panel shows a clear correlation that follows the momentum-transfer scale of the signal event. For $E_R\simeq 248~\mathrm{keV}$ on xenon, $q\equiv|\vec q|=\sqrt{2m_AE_R}\approx 246~\mathrm{MeV}$. For $m_X\gg q$ the mediator propagator reduces to a contact interaction, $\sigma\propto \mu^2 \epsilon^2g_X^2/m_X^4$, so a larger $m_X$ must be compensated by a larger $\epsilon$. This produces the diagonal rise at $\log_{10}(m_X/\mathrm{GeV})\gtrsim -1$. For $m_X\lesssim q$ the propagator scales as $1/q^4$ and the rate becomes independent of $m_X$, producing the flat floor in $\epsilon$ at small $m_X$. The data therefore constrain the combination entering the scattering amplitude, $\epsilon g_X/(m_X^2+q^2)$, rather than $m_X$ and $\epsilon$ separately, which explains the width of the one-dimensional marginals.
  \item The dark gauge coupling is more tightly constrained than $m_X$ and $\epsilon$, with $g_X=0.94^{+0.64}_{-0.43}$. Its $68\%$ interval spans only a factor of about three, and the marginal peaks near $g_X\sim0.5$ with a tail toward the upper edge of the range. This is because $g_X$ is fixed mainly by the requirement of reproducing the observed relic abundance. For annihilation dominated by $\chi_i\chi_i\to XX$, $\langle\sigma v\rangle\propto g_X^4/m_{\rm DM}^2$, so a fixed relic abundance implies $g_X\propto m_{\rm DM}^{1/2}$. This appears in the $(m_{\rm DM},g_X)$ panel as a narrow rising band, with $g_X\simeq0.3$ at $m_{\rm DM}\sim 200~\mathrm{GeV}$ and $g_X\simeq1.8$ at $\sim10~\mathrm{TeV}$. The relic density therefore ties $g_X$ to $m_{\rm DM}$ rather than fixing it independently. Since the inelastic scattering rate scales as $\epsilon^2g_X^2/m_{\rm DM}$, this relation also cancels the explicit mass dependence of the rate along the relic band, which is why the $(m_{\rm DM},\epsilon)$ panel shows no visible correlation.
\end{enumerate}

\section{FOPT and GW signals}\label{sec:fopt}

\begin{figure*}
    \centering
    \includegraphics[width=0.47\linewidth]{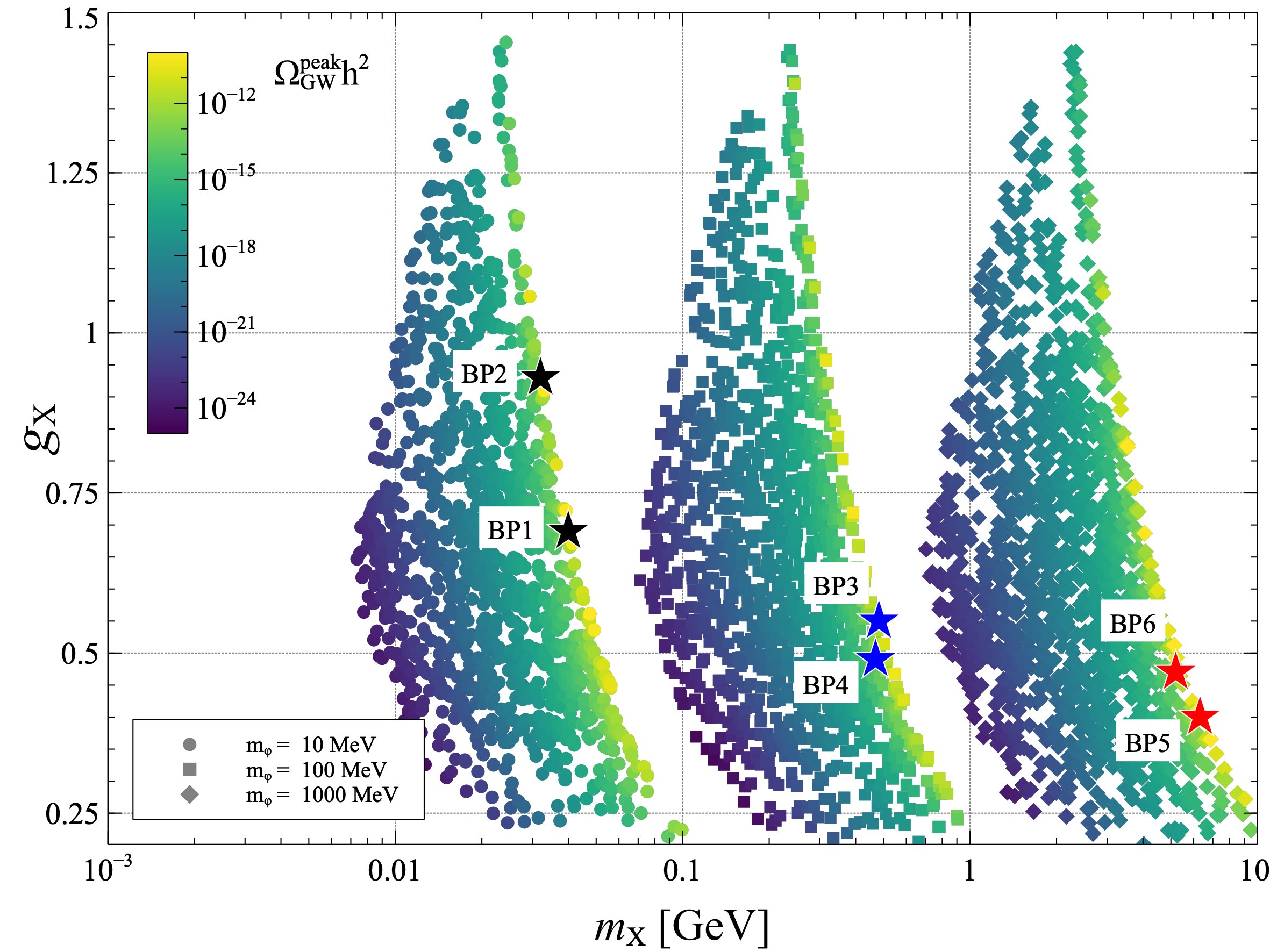}
    \includegraphics[width=0.47\linewidth]{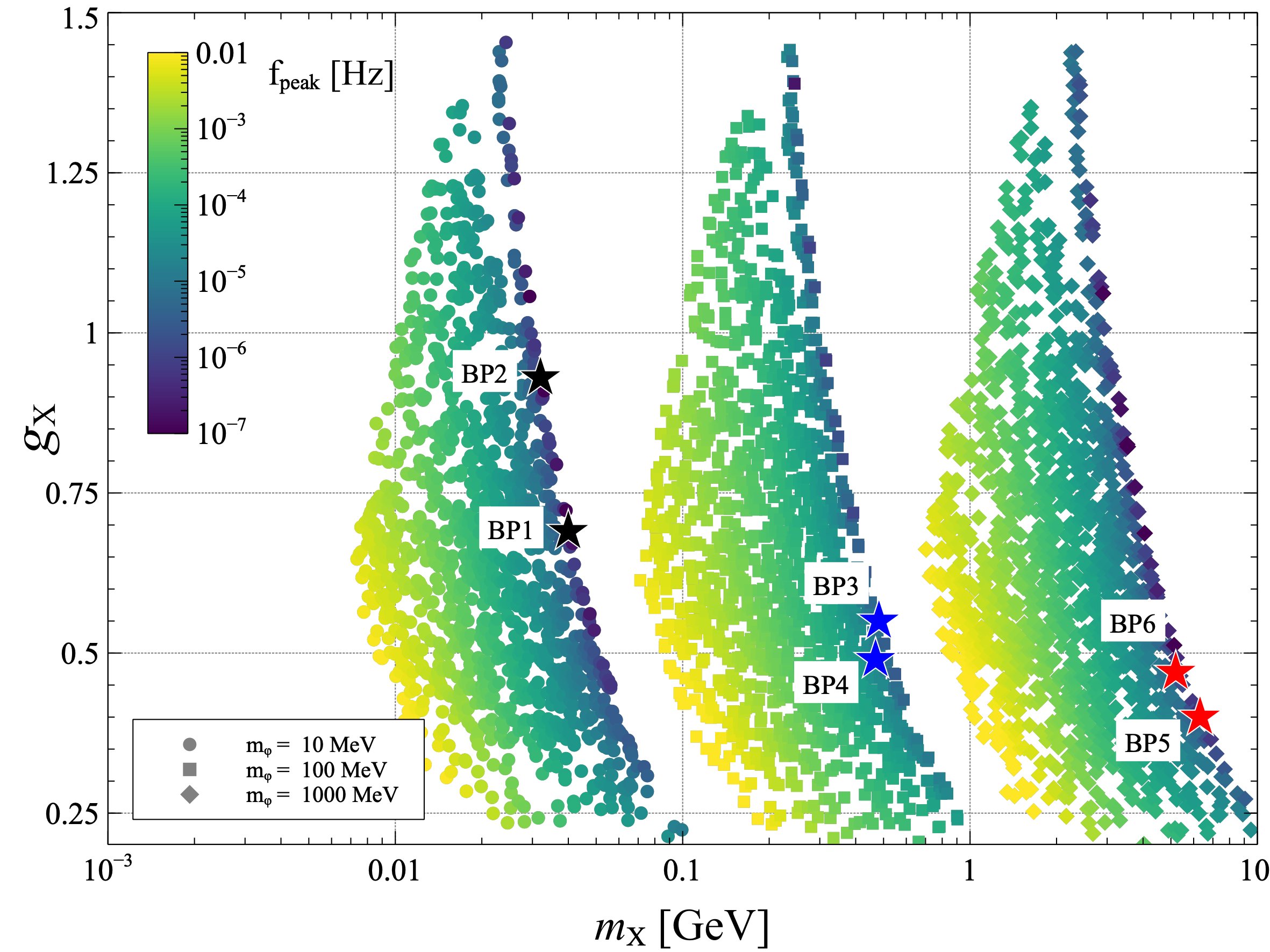}
    \caption{Predicted GW signals from the dark phase transition for samples that simultaneously explain the LZ event and the correct DM relic density, in the $(m_X, g_X)$ plane. The three patches correspond to $m_\phi = 10$ MeV (circles), 100 MeV (squares) and 1000 MeV (diamonds). Left: points colored by the peak amplitude, $\Omega^{\rm peak}_{\rm GW}h^2$. Right: points colored by the frequency, $f_{\rm peak}$ at which the GW amplitude is peaked. Stars denote the benchmark points listed in Table II, whose GW spectra are shown in Fig. 5.}
    \label{fig:gw_signals_1GeV}
\end{figure*}

In this model, the dark $U(1)_X$ symmetry is broken by the vev $v_\phi$ of a dark scalar $\Phi$, which sets both the dark-photon mass, $m_X = 2g_Xv_\phi$, and, via a Yukawa coupling $y$, the pseudo-Dirac splitting, $\delta\sim y\,v_\phi$. These two observables constrain $v_\phi$ very differently. Since $\delta$ depends on the product $y~v_\phi$, the Yukawa coupling can always be retuned to compensate for a change in $v_\phi$, so the LZ-preferred splitting - pinned near its kinematic ceiling $\delta_{\max}(m_{\rm DM})$ (Eq.~\eqref{eq:delta_max}) regardless of $v_\phi$ - carries little information about the scale of symmetry breaking. The dark-photon mass, by contrast, fixes $v_\phi=m_X/2g_X$ directly once $g_X$ is known: $m_X$ is set by the LZ scattering rate, while $g_X$ is fixed largely independently by the relic abundance through secluded annihilation. Because the transition dynamics are controlled by $v_\phi$, the same scale that fixes $m_X$ through $g_X$, the LZ- and relic-preferred region of parameter space makes a corresponding, testable prediction for the GW spectrum, offering an independent cosmological probe of the same dark sector. At tree level, the scalar potential is a simple quartic,
\begin{equation}
    V_0(\phi) = -\tfrac12\mu_\phi^2\phi^2 + \tfrac14\lambda_\phi\phi^4,
    \label{eq:tree_potential}
\end{equation}
parametrized in terms of the physical scalar mass $m_\phi$ (or $m_\phi (T=0)$) and the vev $v_\phi$ via $\mu_\phi^2=m_\phi^2/2$ and $\lambda_\phi=m_\phi^2/2v_\phi^2$, so that $V_0$ alone describes a smooth, barrier-free symmetry breaking. A potential barrier separating the symmetric and broken phases arises only once thermal corrections (see Appendix~\ref{app:fopt_accessories}) are included, and its origin can be traced to the field-dependent boson masses entering the one-loop thermal potential,
\begin{equation}
    m_\phi^2(\phi) = -\mu^2+3\lambda\phi^2, \qquad m_X^2(\phi) = 4g_X^2\phi^2.
\end{equation}

Because $m_X(\phi)\propto\phi$ vanishes identically in the symmetric phase, the dark photon contributes a term to the high-temperature expansion of the thermal potential that scales as $-T\,m_X^3(\phi)\propto -g_X^3T\phi^3$. This negative cubic term is the well-known source of the potential barrier in Abelian Higgs-type transitions. The scalar $\phi$ contributes a smaller, similarly structured cubic term. Because the dark photon carries three polarization degrees of freedom, and its thermal mass grows as $g_X^2\phi^2$, the strength of the barrier, and hence the order and strength of the phase transition, is controlled predominantly by $g_X$.

In a first-order phase transition, the system transitions from a metastable false vacuum to a stable true vacuum via the nucleation of bubbles. The nucleation rate, the probability per unit volume per unit time of forming a critical bubble of true vacuum, is given by
\begin{equation}
    \Gamma (T) \approx T^4 \left( \frac{S_3}{2\pi T} \right)^{3/2} e^{-S_3/T}
\end{equation}
where $S_3(T)$ is the three-dimensional Euclidean action of the critical bubble. The FOPT is characterised by $T_c$, the critical temperature at which the two minima are degenerate and the nucleation temperature, defined by the condition that the nucleation rate in a Hubble volume and a Hubble time reaches $\mathcal{O}(1)$, i.e. $\Gamma~H^{-4}|_{T_n}\approx 1$. A more physically complete criterion is the percolation temperature $T_p$, at which the false-vacuum fraction drops to about $0.71$, so that roughly $30\%$ of the Universe's volume has been converted to the true vacuum and the bubbles form a connected cluster; for the transition to complete, the false-vacuum fraction must then continue to decrease to zero at lower temperatures, which we verify for all benchmark points. We use $T_n$ throughout most of our analysis instead of $T_p$, since for the benchmarks considered here the two temperatures differ only slightly, as is the case in most models with a FOPT that is not strongly supercooled.

\begin{table*}[t]
\centering
\caption{FOPT parameters for the six benchmark points.}
\label{tab:FOPT_BPs}
\renewcommand{\arraystretch}{1.15}
\begin{tabular}{c c c c c c c c c c c c}
\hline\hline
 & $m_\phi\,[\rm MeV]$ & $g_x$ & $m_X\,[\rm MeV]$ & $T_c\,[\rm MeV]$ & $T_n\,[\rm MeV]$ & $\alpha$ &
$\beta/H_{\star}$ & $v_w$ & $m_{\rm DM}\,[\rm GeV]$ & $\delta\,[\rm keV]$ & $\epsilon$\\
\hline
BP1 (\textcolor{black}{\scalebox{1.0}{$\bigstar$}}) & 10 & 0.69 & 40.3 & 11.97 & 6.35 & 0.72 & 127.73 & 0.92 & 1103.66 & 369.03 & $7\times 10^{-5}$\\
BP2 (\textcolor{black}{\scalebox{1.0}{$\bigstar$}}) & 10 & 0.93 & 32 & 9.5 & 6.06 & 0.25 & 283.13 & 0.85 & 2014.03 & 379.79 & $1.0\times 10^{-4}$ \\
BP3 (\textcolor{blue}{\scalebox{1.0}{$\bigstar$}}) & 100 & 0.55 & 482 & 149.84 & 71.16 & 0.83 & 136.9 & 0.92 & 700.69 & 335.05 & $9\times 10^{-6}$\\
BP4 (\textcolor{blue}{\scalebox{1.0}{$\bigstar$}}) & 100 & 0.49 & 469 & 168.55 & 124.1 & 0.10 & 845.7 & 0.77 & 555.98 & 3333.04 & $3.0\times 10^{-5}$\\
BP5 (\textcolor{red}{\scalebox{1.0}{$\bigstar$}}) & 1000 & 0.40 & 6316.9 & 2027.57 & 682.61 & 0.78 & 202.87 & 0.92 & 373.42 & 298.65 & $7\times10^{-4}$\\
BP6 (\textcolor{red}{\scalebox{1.0}{$\bigstar$}}) & 1000 & 0.47 & 5198.11 & 1737.73 & 1019.24 & 0.07 & 414.14 & 0.75 & 511.27 & 318.44 & $6\times 10^{-4}$ \\
\hline\hline
\end{tabular}
\end{table*}

The strength and dynamics of the phase transition are characterized by two dimensionless parameters evaluated at this reference temperature, namely, $\alpha$ and $\beta/H_\star$. Along with the nucleation temperatures, $T_c$ and $T_n$ (see Eqs.~\eqref{eq:alpha} and \eqref{eq:betabyH}), these quantities provide a complete description of the three distinct stochastic gravitational-wave sources.
~\cite{Kamionkowski:1993fg,Caprini:2019egz}: (i) collisions of the expanding bubble walls, (ii) sound waves generated in the plasma after the bubbles collide and merge, and (iii) magnetohydrodynamic turbulence subsequently developing in the plasma. The resulting signal is quantified by the GW energy density per logarithmic frequency interval, normalized to the critical density of the Universe today,
\begin{equation}
    \Omega_{\rm GW}(f) \equiv \frac{1}{\rho_c}\frac{d\rho_{\rm GW}}{d\ln f}, \qquad \rho_c = \frac{3H_0^2}{8\pi G}.
\end{equation}
Because the three sources above act as independent, uncorrelated contributions to the stochastic background, the total signal is well approximated by a linear superposition,
\begin{equation}
    \Omega_{\rm GW}h^2(f) = \Omega_{\rm col}h^2(f) + \Omega_{\rm sw}h^2(f) + \Omega_{\rm turb}h^2(f),
\end{equation}
with each term computed from $\alpha$, $\beta/H_\star$, $v_w$, and $T_n$ using fitting formulae calibrated to hydrodynamic simulations; the explicit spectral templates and coefficients are given in Appendix~\ref{app:gw_accessories}.

\begin{figure}[t]
    \centering
    \includegraphics[width=1.0\linewidth]{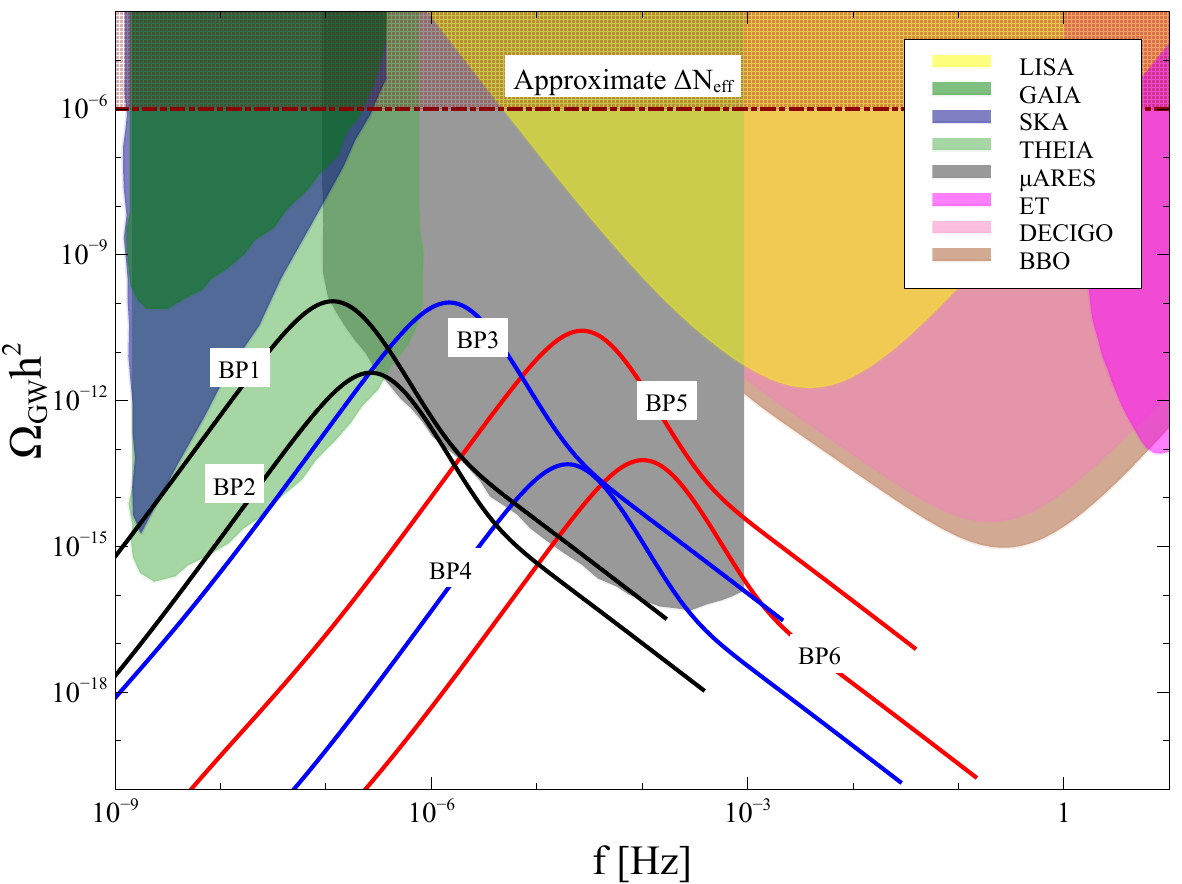}
    \caption{Predicted GW signals from the region of the LZ and Relic density favored parameter space in the plane of frequency vs $\Omega_{\rm GW}h^2$.
    The benchmark points BP1--BP6 are also shown here in star shaped symbols. The shaded regions show the projected sensitivities of BBO, ET, LISA, GAIA, THEIA, DECIGO, SKA, and $\mu$ARES. The dot-dashed horizontal line denotes the approximate upper bound on $\Delta N_{\rm eff}$ from Planck 2018 data.}
    \label{fig:GW_spectrum}
\end{figure}

Fig.~\ref{fig:gw_signals_1GeV} shows the peak GW amplitude (left panel) and the corresponding peak frequency (right panel) of the predicted GW signal in the $m_X$–$g_X$ plane, highlighting the parameter space favored by the combined analysis of the LZ event and the DM relic density. Although the mass of the scalar field responsible for the symmetry breaking, and hence for the GW signal, does not enter the likelihood analysis, its value is crucial for the FOPT analysis. For a given $m_\phi$, say $10~{\rm MeV}$, the VEV is fixed via $v_\phi = m_\phi/\sqrt{2\lambda_\phi}$. For values of the quartic self-coupling $\lambda_\phi$ within the range allowed by perturbativity and boundedness from below, the resulting $v_\phi$ is mapped onto combinations of $m_X$ and $g_X$ through the gauge boson mass relation. The corresponding region is the leftmost patch in the left panel of Fig.~\ref{fig:gw_signals_1GeV}. Likewise, $m_\phi=100~{\rm MeV}$ and $m_\phi=1000~{\rm MeV}$ select two further patches from the same set of parameters sampled by the LZ and relic density likelihood, shown as the middle and rightmost regions of the left panel, respectively. Within each patch, the GW amplitude grows with increasing $m_X$, with the peak amplitude ranging from $10^{-24}$ up to $10^{-10}$. Conversely, in the right panel figure, the peak frequency decreases as $m_X$ increases, placing the detectable signals in the range from the mHz down to the $\mu$Hz band and below. This anticorrelation, whereby larger amplitudes appear at lower frequencies, is closely tied to the FOPT dynamics. For a sound-wave source, the peak frequency and amplitude are controlled by $\alpha$, $\beta/H_\star$, $T_\star$ and $v_w$. As detailed in Appendix~\ref{app:gw_accessories}, the sound-wave peak frequency scales as $f_{\rm sw}\propto(\beta/H_\star)\,T_\star$, so a larger $T_\star$ or $\beta/H_\star$ shifts the spectrum to higher frequencies. The sound-wave amplitude, however, scales as $\Omega_{\rm sw}\propto(\beta/H_\star)^{-1}$ and is therefore suppressed for large $\beta/H_\star$. Consequently, different regions of the viable parameter space yield distinct GW signals. We select six benchmark points (BPs), marked by black, blue and red stars, whose parameter values are listed in Table~\ref{tab:FOPT_BPs}. For three representative cases, BP1, BP3 and BP6, Fig.~\ref{fig:likelihood} shows the $1\sigma$ region preferred by the LZ event, with (solid) and without (dashed) the relic density constraint imposed. In all three cases, for a fixed value of $g_X$, the relic density requirement strongly restricts the allowed DM mass, but leaves the mediator mass $m_X$ largely unconstrained. The viable range of $m_X$ is therefore set by the combined LZ and relic density region (the solid contour). Since the GW signal depends on $m_X$ through the scale of symmetry breaking, different values of $m_X$ within this region yield different GW spectra, as can be seen directly from Fig.~\ref{fig:gw_signals_1GeV}.

Fig.~\ref{fig:GW_spectrum} shows the projected GW sensitivity regions of BBO~\cite{Yagi:2011wg,Schmitz:2020syl}, ET~\cite{Punturo:2010zz}, LISA~\cite{LISA:2017pwj,LISACosmologyWorkingGroup:2022kbp,Schmitz:2020syl}, GAIA~\cite{Garcia-Bellido:2021zgu}, THEIA~\cite{Garcia-Bellido:2021zgu}, DECIGO~\cite{Seto:2001qf,Yagi:2011wg,Schmitz:2020syl}, SKA~\cite{Weltman:2018zrl}, and $\mu$ARES~\cite{Sesana:2019vho}, overlaid with the GW signals from the parameter space favored by the DM relic density and the LZ event, for $m_\phi=10$ MeV (black), 100 MeV (blue) and 1000 MeV (red). The dot-dashed horizontal line at $\Omega_{\rm GW}h^2\sim10^{-6}$ indicates the approximate upper bound on the GW energy density from $\Delta N_{\rm eff}$, derived from Planck 2018 data~\cite{Planck:2018vyg}. Each spectrum rises as $f^{3}$ below the peak. Near the peak, sound waves dominate and the spectrum falls steeply as $f^{-4}$ above $f_{\rm sw}$. At higher frequencies, the slower turbulence tail ($\propto f^{-5/3}$) overtakes the sound-wave contribution and sets the right tail. The peak frequencies span roughly $10^{-7}$--$10^{-4}$~Hz, so the signals lie in the $\mu$Hz--nHz window probed by THEIA, $\mu$ARES and SKA, while remaining below the sensitivity of LISA, DECIGO and BBO. Since the peak frequency grows with $T_\star$ and $\beta/H_\star$ while the amplitude decreases with $\beta/H_\star$, reaching the mHz band or higher requires a larger $T_\star$ together with a sufficiently large $\alpha$, which is not realized here. A larger $T_\star$ with reasonable $\alpha$ and $\beta/H_{\star}$ would require the symmetry-breaking scale to be at a much higher temperature. This implies a much heavier mediator, which, though favored by LZ, is ruled out by terrestrial searches (see Fig.~\ref{fig:eps_MX}). Supercooling could also enhance the peak amplitude, but strong supercooling dumps a large amount of vacuum energy into the SM plasma and reheats the Universe at the end of the transition. If the DM relic density is set during or after this period, it would be substantially diluted~\cite{Huang:2026qdc}, an effect we do not include in our analysis. Hence, all the observable spectra lie at least two orders of magnitude below the $\Delta N_{\rm eff}$ bound, consistent with Planck 2018, and can only be probed by future astrometric ( THEIA) and $\mu$Hz-band ($\mu$ARES) detectors.

\section{Experimental Constraints}\label{sec:expt_constraints}

\begin{figure}
    \centering
    \includegraphics[width=1.\linewidth]{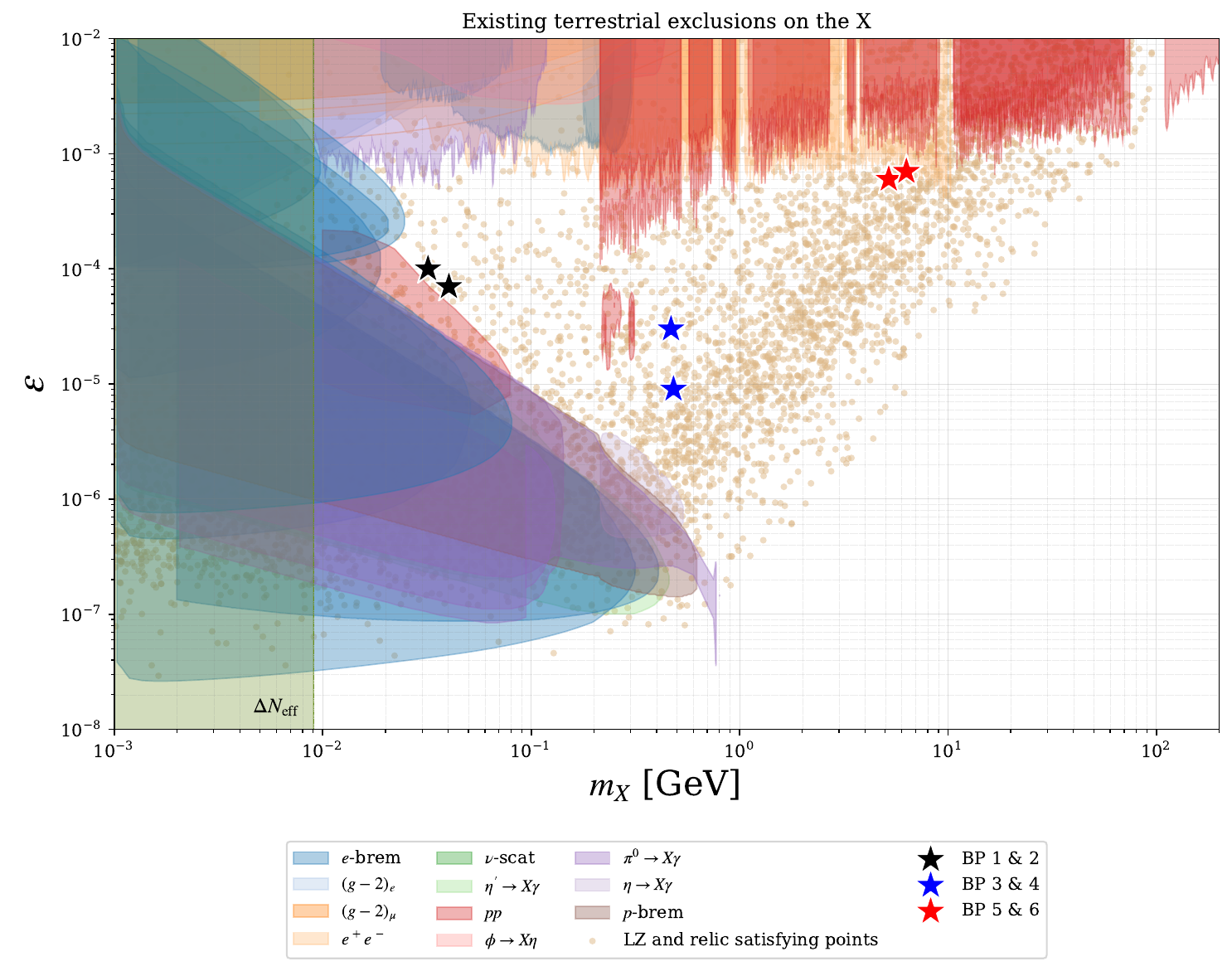}
    \caption{Existing terrestrial exclusions against the viable parameter space for both LZ event and correct relic are projected in the plane of $(m_X,\epsilon)$ by using \texttt{DarkCast} package. The BPs from Table~\ref{tab:FOPT_BPs} are shown by colored stars.}
    \label{fig:eps_MX}
\end{figure}

We project the existing constraints from terrestrial searches to the ($m_X,\epsilon$) plane using the \texttt{DarkCast} package \cite{Ilten:2018crw}. These include beam-dump experiments (E137, E141, E774, CHARM, NuCAL, NOMAD, PS191, Orsay)~\cite{Andreas:2012mt,Riordan:1987aw,Tsai:2019buq,NOMAD:2001eyx,Bernardi:1985ny,Davier:1989wz}, collider and fixed-target searches (LHCb, BaBar, KLOE, NA48, NA62, NA64, BESIII, CMS, APEX, HPS, A1)~\cite{LHCb:2017trq,LHCb:2019vmc,BaBar:2016sci,KLOE-2:2018kqf,NA482:2015wmo,NA62:2023qyn,BESIII:2017fwv,CMS:2019kiy,APEX:2011dww,HPS:2018xkw,A1:2011yso,Merkel:2014avp}, and precision measurements of the electron~\cite{Bodas:2021fsy} and muon~\cite{Muong-2:2025xyk} anomalous magnetic moments. We recast these existing dark-photon exclusions directly onto our $X$ boson, since it couples to SM fermions via kinetic mixing as given in Eq.~\eqref{eq:Xcoupling}. Apart from that, having a lighter mediating particle with coupling to SM fermion can affect the dynamics during the BBN epoch, which can be parametrized using $\Delta N_\text{eff}$. Therefore, we provide a conservative bound on $m_X\gtrsim 9$ MeV which is excluded by the Planck 2018 at 2$\sigma$~\cite{Ibe:2019gpv}. The resulting bounds are casted by colored shaded regions in Fig.~\ref{fig:eps_MX}. We then project our model parameter space that explains both the LZ event and observed DM relic density constraint using scattered points against the existing above discussed constraints to show the viable region. Moreover, the BBN bound is also imposed on the lifetime of $X$ by keeping $\tau_{m_X}<1~$s, thus excluding smaller $\epsilon< 10^{-10}$, which lie outside our explored parameter space. We find that most of our parameter space still remain viable for the future probes.

\section{Conclusion}\label{sec:conclusion}

In this work, we have investigated inelastic dark matter scattering within a generic $U(1)_X$ extension of the SM, containing a scalar field $\phi$ and a fermion $\chi$ charged under the new gauge symmetry. Spontaneous symmetry breaking, induced by the scalar vacuum expectation value $v_\phi$, generates both the mass of the gauge mediator $X^\mu$ and small Majorana mass terms for the fermion. In the pseudo-Dirac limit, these terms split the fermion into two nearly degenerate Majorana mass eigenstates, $\chi_1$ and $\chi_2$, with a mass splitting $\delta \equiv m_{2}-m_{1}\propto yv_\phi$.
The off-diagonal gauge interaction connects the two states, while a small gauge kinetic mixing $\epsilon$ provides a portal to the SM. This structure realizes endothermic scattering, $\chi_1 N\rightarrow\chi_2 N$, in which part of the incident dark matter kinetic energy is converted into the excitation energy $\delta$.

We examined whether this framework can accommodate the reported LZ recoil candidate at $E_R\simeq248~\mathrm{keV}$ while remaining consistent with the absence of events at lower energies. The leading interaction is the proton-dominated, spin-independent operator $\mathcal{O}_1$, with recoil spectra computed using \texttt{WimpyDD} for a $2.84~$tonne-yr exposure. The inelastic threshold provides the mechanism for suppressing low-energy recoils. Since direct detection and thermal freeze-out depend on overlapping model parameters, we performed a combined likelihood analysis of the LZ event counts and the observed relic abundance, using static nested sampling to obtain marginalized posteriors and $68\%$ and $95\%$ credible regions.

The posterior for the dark matter mass is only weakly constrained from below and turns over in the multi-TeV regime before the upper prior boundary of $20~\mathrm{TeV}$. The mass splitting $\delta$ is sharply asymmetric, favoring relatively large values below the kinematic ceiling set by $v_{\mathrm{max}}$, which reflects the competition between suppressing low-energy recoils and retaining enough kinematically accessible dark matter to produce the high-energy event. The mediator mass $m_X$ spans the sub-GeV to few-GeV range, and the kinetic mixing has a median of $\epsilon \sim 10^{-6}$ with a $68\%$ credible interval covering roughly four orders of magnitude. The dark gauge coupling, fixed mainly by the relic abundance, is $g_X = 0.965^{+0.631}_{-0.431}$. 

Because $m_X$ and $\delta$ share a common origin in $U(1)_X$ breaking, these posteriors constrain $v_\phi$ and the Yukawa coupling $y$, linking the direct-detection and relic-density results to the scalar sector and motivating the study of a gravitational-wave signal from a first-order phase transition. The GW prediction nevertheless depends on the scalar potential and is not uniquely fixed by the recoil candidate and relic abundance alone. For the benchmark scalar masses
$m_\phi=(0.01,\,0.1,\,1)~\mathrm{GeV}$, we explored the GW spectra associated with the $(m_X,g_X)$ regions supported by the posterior analysis. The largest predicted peak amplitudes reach approximately $10^{-9}$, with the peaks occurring below the $\mu\mathrm{Hz}$ band and substantially weaker emission at higher frequencies. Consequently,
observational prospects depend on the overlap of the full predicted spectra with the frequency-dependent sensitivities of the relevant experiments. These low-frequency signals motivate comparisons with pulsar-timing observations, including those enabled by SKA, astrometric measurements with GAIA and the proposed THEIA mission, and the proposed $\mu$ARES observatory. The GW frequency in the sensitivity range of the LISA observatory and higher frequency based observatories is strongly disfavored by the combined LZ event and the terrestrial search bounds.

We have also confronted the favored parameter space with terrestrial and cosmological constraints. Beam-dump, fixed-target, and collider searches restrict the
$(m_X,\epsilon)$ plane, while bounds on
$\Delta N_{\mathrm{eff}}$ constrains the change in the SM neutrino temperature after its decoupling, due to the presence of the $X$ boson. Portions of the model parameter space satisfying both the LZ and relic density constraint remain compatible with all the existing bounds considered in this work, leaving viable parameter space for further experimental tests.

Our results establish a viable connection between endothermic dark matter scattering, thermal relic production, and the cosmological dynamics of a broken dark gauge symmetry. The LZ recoil candidate serves as a phenomenological target rather than evidence for a dark matter signal. Within this interpretation, the surviving parameter space motivates complementary tests through future recoil measurements, searches for light vector mediators, and low-frequency gravitational-wave observations.
\appendix
\section{NREFT approach for Direct Detection}\label{app:nreft}
The kinetic mixing between the dark $U(1)_X$ gauge boson and the hypercharge gauge boson induces, after electroweak symmetry breaking and diagonalization of the neutral gauge sector, a coupling of the physical dark gauge boson to the electromagnetic current. At leading order in the kinetic-mixing parameter, the interaction relevant for direct detection can be written as
\begin{equation}
    \mathcal{L}_{X{\rm EM}}= -\epsilon e c_W X_\mu J_{\rm EM}^{\mu},
\end{equation}
where $e$ is the electromagnetic coupling and
$c_W \equiv \cos\theta_W$. The off-diagonal dark-sector interaction is
\begin{equation}
    \mathcal{L}_{\chi X}
    =i g_{12} X_\mu \overline{\chi}_2\gamma^\mu \chi_1,
    \qquad
    g_{12}=g_X\sin 2\theta,
\end{equation}
where $g_{12}=g_X$ for the maximal mixing of our Pseudo Dirac case. At the quark level, the scattering amplitude is given by 
\begin{equation}
    \mathcal{M} \propto \frac{g_X\,\epsilon e c_W}{q^2+m_X^2}
    (\overline{\chi}_2\gamma^\mu \chi_1)
    \langle N|J_{\rm EM}^{\mu}|N\rangle.
\end{equation}

For an on-shell nucleon, the matrix element of the electromagnetic current is conventionally parametrized in terms of the Dirac and
Pauli electromagnetic form factors $F_1^N$ and $F_2^N$ as \cite{Lin:2021umz,Alberico:2008sz}
\begin{equation}
    \begin{split}
    \langle N(k')|J_{\rm EM}^{\mu}|N(k)\rangle
    =&\overline{u}_N(k')
    \left[
        F_1^N(q^2)\gamma^\mu \right.\\   
      & \left. +\frac{i F_2^N(q^2)}{2m_N}
        \sigma^{\mu\nu}q_\nu
    \right]
    u_N(k),
    \end{split}
    \label{eq:nucleon_EM_current}
\end{equation}
where
\begin{equation}
    q^\mu=k^\mu-k'^\mu=p'^\mu-p^\mu,
    \qquad
     -q^2>0 .
\end{equation}
Here $F_1^N$ and $F_2^N$ denote the Dirac and Pauli form factors, respectively. Ignoring the Pauli form factors, their normalization at vanishing momentum transfer is fixed by the electric charges and anomalous magnetic moments of the nucleons,
\begin{align}
    F_1^p(0) &= 1, & F_1^n(0) &= 0,
    \label{eq:F1_normalization}
\end{align}

Keeping only the leading Dirac contribution in
Eq.~\eqref{eq:nucleon_EM_current}, the vector--vector scattering amplitude mediated by $X$ is
\begin{equation}
\label{eq:M_VV_EM}
    \mathcal{M}_{VV}^{N}
    \simeq
    \frac{
        g_{12}\,\epsilon e c_W
    }{
        q^2+m_X^2
    }
    F_1^N(q^2)
    \left(
        \overline{u}_2\gamma^\mu u_1
    \right)
    \left(
        \overline{u}_N'\gamma_\mu u_N
    \right),
\end{equation}
up to an overall sign convention that is irrelevant for the single-amplitude rate.

Following the standard nonrelativistic effective-field-theory normalization \cite{Anand:2013yka,Barello:2014uda}, we define
\begin{equation}
    \mathcal{M}_{\rm NR}=
    \frac{\mathcal{M}_{\rm rel}}
         {4m_\text{DM} m_N}
    =\sum_i c_i^N \mathcal{O}_i^N .
\end{equation}
For the vector--vector bilinear, the leading nonrelativistic
reduction is
\begin{equation}
    \frac{1}{4m_\text{DM} m_N}
    \left( \overline{u}_2\gamma^\mu u_1 \right)
    \left( \overline{u}_N'\gamma_\mu u_N \right)
    = \mathcal{O}_1,
\end{equation}
where
\begin{equation}
    \mathcal{O}_1=1_\chi 1_N .
\end{equation}
The extension of the standard nonrelativistic operator formalism to inelastic dark-matter transitions is discussed in Ref.~\cite{Barello:2014uda}. It follows from Eq.~\eqref{eq:M_VV_EM} that the leading electromagnetic contribution to the coefficient of $\mathcal{O}_1$ is
\begin{equation}
    c_1^N(q^2)
    =\frac{g_X\sin 2\theta\,\epsilon e c_W}{q^2+m_X^2}
    F_1^N(q^2).
\label{eq:c1_general}
\end{equation}

In the zero-momentum, or leading charge, approximation,
Eq.~\eqref{eq:F1_normalization} therefore gives
\begin{align}
c_1^p(q^2)&\simeq \frac{g_X\sin 2\theta\,\epsilon e c_W}{q^2+m_X^2},
\label{eq:c1p}
\\[2mm]
c_1^n(q^2)&\simeq0.
\label{eq:c1n}
\end{align}

The neutron coefficient, and the $Z$-induced correction to $c_1^p$, are suppressed by $1/m_Z^2$. For a light mediator we keep the propagator momentum-dependent, which does not alter the nuclear response, as noted in Appendix A.3 of \cite{Anand:2013yka}

\section{Effective Potential for FOPT}\label{app:fopt_accessories}
The one-loop finite-temperature effective potential $V_{\rm eff}(\phi,T)$ governing the dark scalar $\phi$ is given by 
\begin{equation}
    \begin{split}
        V_{\rm eff}(\phi,T) =& V_0(\phi) + V_{\rm CW}(\phi) + V_{\rm ct}(\phi)\\
        &+ V_T(\phi,T) + V_{\rm daisy}(\phi,T),
    \end{split}
\end{equation}
Here, $V_0$ denotes the tree level potential described in Eq.~\eqref{eq:tree_potential}, and $V_{CW}$ is the Coleman-Weinberg potential \cite{Coleman:1973jx}, the zero-temperature quantum correction obtained by integrating out fluctuations of each field around the background value $\phi$,
\begin{equation}
    V_{\rm CW}(\phi) = \sum_i \frac{n_i}{64\pi^2}\,m_i^4(\phi)\left[\ln\frac{m_i^2(\phi)}{\mu_R^2}-c_i\right],
\end{equation}
where the sum runs over all bosonic species with field-dependent mass $m_i(\phi)$ and degeneracy $n_i$, $\mu_R$ is the renormalization scale, and $c_i$ is a scheme-dependent constant ($c_i=5/6$ for gauge bosons, $3/2$ for scalars, in $\overline{\rm MS}$-like schemes). This term encodes the leading quantum corrections to the tree-level shape of the potential. Because $V_{\rm CW}$ generically shifts the location and curvature of the minimum away from the tree-level values, a set of counterterm potential, $V_{\rm ct}$ is added,
\begin{equation}
    V_{\rm ct}(\phi) = \tfrac12\delta\mu^2\phi^2 + \tfrac14\delta\lambda\,\phi^4,
\end{equation}
with $\delta\mu^2$ and $\delta\lambda$ fixed by the renormalization conditions that the physical scalar mass and vev remain exactly $m_\phi$ and $v_\phi$ after including the one-loop correction. This ensures that the input parameters of the model retain their physical meaning at one loop. Finite-temperature (thermal) potential \cite{Dolan:1973qd,Quiros:1999jp}, $V_T(\phi,T)$ term captures the effect of the thermal bath on the potential,
\begin{equation}
    V_T(\phi,T) = \frac{T^4}{2\pi^2}\sum_i n_i\,J_{B/F}\!\left(\frac{m_i^2(\phi)}{T^2}\right),
\end{equation}
with $J_B$ ($J_F$) the standard bosonic (fermionic) thermal functions, given by 
\begin{equation}
    J_{B/F} (y^2)=\int_0^\infty x^2 {\rm log}\left(1\mp e^{-\sqrt{x^2+y^2}}\right)~dx
\end{equation}

At high temperature, an infinite series of higher-loop ``daisy" diagrams becomes comparable in size to the one-loop thermal potential and must be resummed. This is implemented by shifting each boson's field-dependent mass by its thermal (Debye) mass, $m_i^2(\phi)\to m_i^2(\phi)+\Pi_i(T)$, before substituting into the leading high-temperature (cubic) term,
\begin{equation}
    V_{\rm daisy}(\phi,T) = -\frac{T}{12\pi}\sum_i n_i\left[\left(m_i^2(\phi)+\Pi_i(T)\right)^{3/2} - m_i^3(\phi)\right].
\end{equation}

The Debye mass $\Pi_i(T)\propto T^2$ is the effective mass a particle acquires from its interactions with the surrounding thermal plasma, analogous to electric charge screening in an ordinary plasma; following the Arnold–Espinosa prescription, we apply this resummation to the scalar and to the longitudinal gauge-boson polarization, while the transverse gauge polarizations, which are not subject to the same infrared enhancement, retain their unresummed cubic contribution. 

The strength and dynamics of the transition are characterized by two dimensionless parameters computed at this reference temperature - namely, the transition strength and the inverse duration of the transition, denoted by $\alpha$ and $\beta/H_\star$, where $H_\star = H(T_n)$. The transition strength,
\begin{equation}
    \alpha = \frac{1}{\rho_{\rm rad}}\left[\Delta V - \frac{T}{4}\frac{d\Delta V}{dT}\right]_{T=T_n},
    \label{eq:alpha}
\end{equation}
compares the latent heat released in the transition to the radiation energy density $\rho_{\rm rad}$ of the plasma at $T_n$, with $\Delta V$ the free-energy difference between the false and true vacua. The inverse duration,
\begin{equation}
    \frac{\beta}{H_\star} = T_n\,\frac{d}{dT}\left(\frac{S_3}{T}\right)\bigg|_{T=T_n},
    \label{eq:betabyH}
\end{equation}
measures how rapidly the bounce action falls near the nucleation temperature, and hence how quickly the transition completes relative to the Hubble time.

\section{Gravitational-Wave Signal from FOPT}\label{app:gw_accessories}
The stochastic gravitational-wave spectrum generated by a first-order phase transition receives contributions from bubble collisions, sound waves in the plasma, and magnetohydrodynamic turbulence. The total GW spectrum is therefore written as the sum of the individual contributions,
\begin{equation}
    \Omega_{\rm GW}(f)=\Omega_{\rm col}(f)+\Omega_{\rm sw}(f)+\Omega_{\rm turb}(f),
\end{equation}
whose amplitudes and spectral shapes are determined by the transition parameters and the corresponding energy-transfer efficiencies. Denote $B=\beta/H_\star$, $K_i=\frac{\kappa_i \alpha}{1+\alpha}$, $\mathcal{G}=h^2\Omega_{\rm GW}$ and
$\mathcal{R}=\frac{T_n}{100~\rm GeV}\left( \frac{g_s(T_n)}{100} \right)^{1/6}$, we have the following
\\
\\
\textit{Bubble Collisions :}
\begin{equation*}
    \begin{split}
        \mathcal{G}_{\rm col}(f) =&1.67\times10^{-5}B^{-2}K_\phi^2 \left( \frac{100}{g_\star} \right)^{1/3}\frac{0.11 v_w^3}{0.42 +v_w^2}S_{\rm col}(x)\\
       & {\rm with~} S_{\rm col}(x)=\frac{3.8~x^{2.8}}{1+2.8~x^{3.8}};\quad x=\frac{f}{\hat{f}_{\rm 
       col}}\\
       &\hat{f}_{\rm col}=1.65 \times 10^{-5} {\rm Hz}~B\mathcal{R} \frac{0.62}{1.8-0.1v_w+v_w^2}.
    \end{split}
\end{equation*}
\\
\textit{Sound Waves :}
\begin{equation*}
    \begin{split}
        \mathcal{G}_{\rm sw}(f) =&2.65\times10^{-6}B^{-1}K_{\rm sw}^2 \left( \frac{100}{g_\star} \right)^{1/3}v_w~S_{\rm sw}(x) \Upsilon\\
       & {\rm with~} S_{\rm sw}(x)=x^3\left(\frac{7}{4+3x^2}\right)^{7/2};\quad x=\frac{f}{\hat{f}_{\rm 
       sw}}\\
       &\hat{f}_{\rm sw}=1.9 \times 10^{-5} {\rm Hz}\frac{B\mathcal{R} }{v_w}.
    \end{split}
\end{equation*}
\\
\textit{Turbulence :}
\begin{equation*}
    \begin{split}
        \mathcal{G}_{\rm turb}(f) =&3.35\times10^{-4}B^{-1}K_{\rm turb}^{3/2} \left( \frac{100}{g_\star} \right)^{1/3}v_w~S_{\rm turb}(f)\\
       & {\rm with~}\\
       &S_{\rm turb}(f)=\frac{(f/\hat{f}_{\rm turb})^3}{(1+f/\hat{f}_{\rm turb})^{11/3}(1+8\pi f/h_\star)},\\
       &\hat{f}_{\rm turb}=2.7 \times 10^{-5} {\rm Hz}\frac{B\mathcal{R} }{v_w},~{\rm and}\\
       &h_\star=1.65\times10^{-5} {\rm Hz}~\mathcal{R}.
    \end{split}
\end{equation*}
The factor $\Upsilon=1-\frac{1}{\sqrt{1+2H_\star \tau_{\rm sw}}}$, accounts for the finite lifetime of the sound-wave period. The duration of the sound-wave source is estimated as $H_\star \tau_{\rm sw} \simeq \frac{H_\star R_\star}{\overline{U}_f}$, where the average bubble separation is $H_\star R_\star=(8\pi)^{1/3} \frac{v_w}{B}$, and the root-mean-square fluid velocity is $\overline{U}^2_f \simeq\frac{3}{4}K_{\rm sw}$ \cite{Guo:2020grp}. 

For a terminal Jouguet wall assumption, the consistent leading approximation \cite{Espinosa:2010hh} is
\begin{equation}
    \kappa_{\rm col} \approx 0,~v_w=v_J(\alpha),~k_{\rm sw}=\frac{\sqrt
    \alpha}{0.135+\sqrt{0.98+\alpha}}
    \label{eq:Jouguet_approx}
\end{equation}
where $v_J(\alpha)=\frac{1/\sqrt{
3}+\sqrt{\alpha^2+2\alpha/3}}{1+\alpha}$ and for the turbulence efficiency factor, we adopt the commonly used approximation, $\kappa_{\rm turb}\simeq 0.1 \kappa_{\rm sw}$. The total spectrum is the sum of the sound-wave and turbulence contributions, $\Omega_{\rm GW}h^2\simeq\Omega_{\rm sw}h^2+\Omega_{\rm turb}h^2$, and its shape reflects the two sources.


%

\end{document}